\documentclass[conference]{IEEEtran}
\usepackage{mathtools}
\usepackage{listings}
\usepackage{xcolor}
\usepackage{dblfloatfix}

\usepackage{float}
\usepackage{mdwlist}
\usepackage{paralist}
\usepackage{graphics}
\usepackage{enumerate}
\usepackage{dblfloatfix}

\DeclareMathSizes{9}{19}{13}{9}
\usepackage[top=1in, bottom=1in, left=1in, right=1in]{geometry}

\usepackage[tight,footnotesize]{subfigure}

\usepackage{booktabs}
\usepackage{lipsum}
\usepackage{tabularx}

\usepackage{placeins}
\newcommand{\WordCapacity}{capacity}

\usepackage{caption}
\newcommand{\WordMemory}{memory}

\newcommand{\MemoryCapacity}{C_{\WordMemory}}

\newcommand{\WordDataBlock}{data block }

\newcommand{\NumberBlock}{n_{block}}
\newcommand{\DataBlockCapacity}{C_{\WordCapacity}}

\newcommand{\Associativity}{associativity }
\newcommand{\NumberSet}{n_{set}}

\newcommand{\SliceID}{slice\_id }

\newcommand{\WordSet}{set}
\newcommand{\SetIndex}{set\_index}
\newcommand{\NumberSetEachSlice}{n_{\WordSet}}

\newcommand{\PhysicalAddressAbbreviation}{pa}

\newcommand{\MA}{\PhysicalAddressAbbreviation}
\newcommand{\MATOSliceID}{ma\_to\_slice\_id}
\newcommand{\MATOSetIndex}{ma\_to\_set\_index}

\newcommand{\PhysicalAddressWidth}{n}

\newcommand{\NUMSLICE}{n_{slice}}
\newcommand{\NUMSET}{n_{set}}

\newcommand{\WIDTHSetIndex}{$n\_slice$}

\newcommand{\SandyBridge}{Sandy Bridge }

\usepackage{etoolbox}
\makeatletter
\patchcmd{\@makecaption}
  {\scshape}
  {}
  {}
  {}
\makeatother

\newcommand{\otoprule}{\midrule[\heavyrulewidth]}

\ifCLASSINFOpdf
\else
\fi
\begin{document}
%
\title{\huge{\textbf{Cracking Intel Sandy Bridge's Cache Hash Function}}}

\author{


\IEEEauthorblockN{Zhipeng Wei, Zehan Cui, Mingyu Chen}
\IEEEauthorblockA{\\Insitute of Computing Technology, Chinese Academy of Sciences \\ $\left\{weizhipeng, cuizehan, cmy\right\}$@ict.ac.cn}
}

%


\maketitle

\begin{abstract}
On Intel Sandy Bridge processor, last level cache (LLC) is divided into cache slices and all physical addresses are distributed across the cache slices using an hash function. 
With this undocumented hash function existing, it is impossible to implement cache partition based on page coloring.
This article cracks the hash functions on two types of Intel Sandy processors by converting the problem of cracking the hash function to the problem of classifying data blocks into different groups based on eviction relationship existing between data blocks that are mapped to the same cache set.
Based on the cracking result, this article proves that it's possible to implement cache partition based on page coloring on cache indexed by hashing.
\end{abstract}


%
\IEEEpeerreviewmaketitle

\section{Introduction}
\begin{figure}[!htbp]
\centering
\includegraphics[width=0.8\columnwidth]{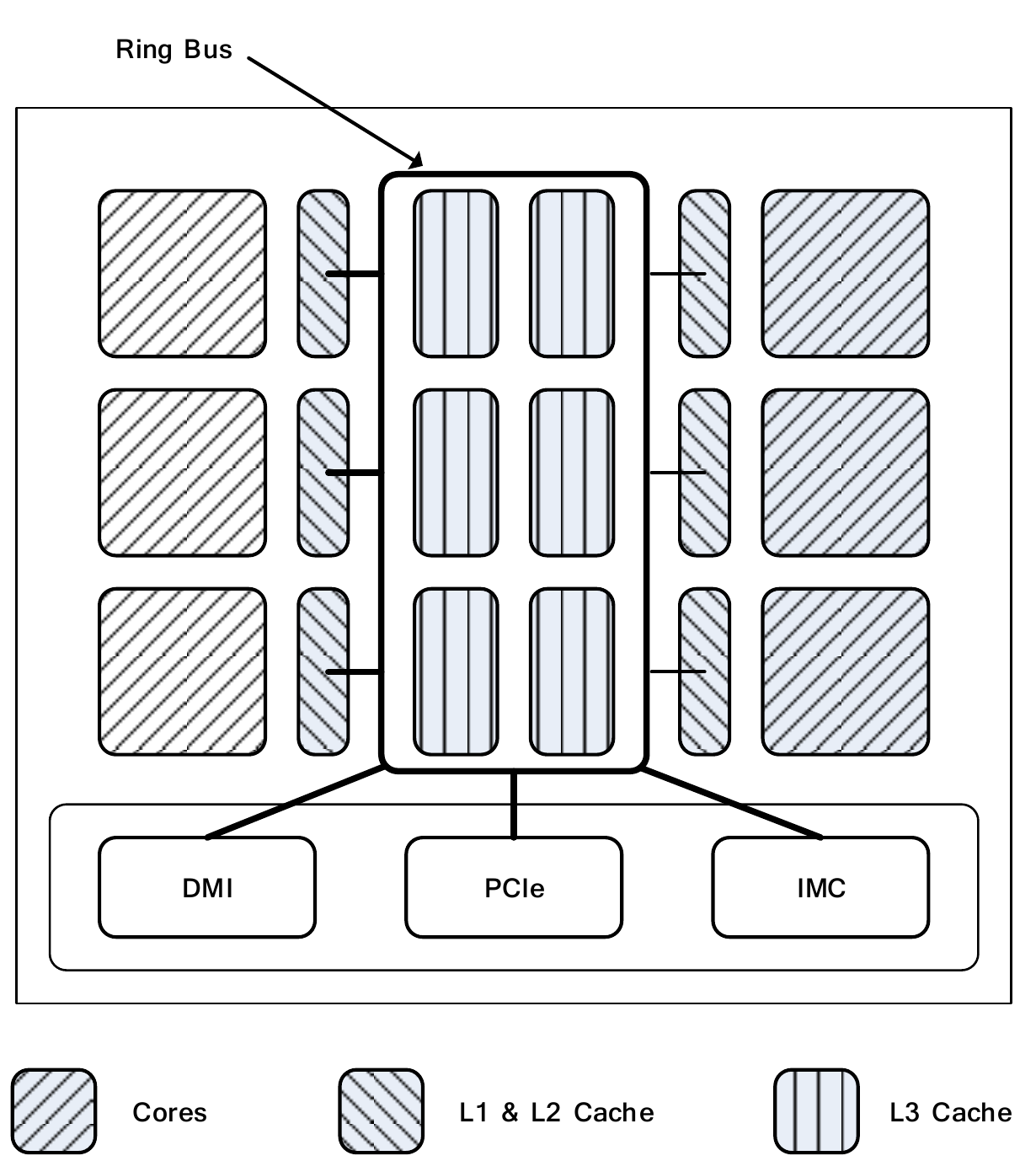}
\caption{LLC organization on Intel Sandy processor.}
\label{fig:sixslices}
\end{figure}

Cache plays an important role in bridging the gap between the speed of processor and main memory. Many cache architectures have been proposed in history. Hash is an important technique to improve cache performance, such as hash-rehash cache. As the succeeding generation to Nehalem, one of Sandy Bridge processor's new features is that LLC is divided into several slices which are connected by a ring bus, as shown in Figure~\ref{fig:sixslices}. And the location of a given data block on LLC is decided by an undocumented hash function.  

This article proposes a novel method using HMTT~\cite{bao2008hmtt} to crack the hash function and further verifies the correctness of the cracking result based on the phenomenon that when the number of accessed data blocks that are mapped to the same cache set exceeds the associativity of cache set, average access latency increases sharply. Compared to the statement that page coloring doesn't work on caches that are indexed using hashing~\cite{sanchez2012scalable}, this articles ported User Level Cache Control, which is a software runtime library used to improve cache performance by implementing cache partition based on page coloring, and proves hat it is possible to implement cache partition using page coloring on Intel Sandy Bridge processors. 

This articles has the following contributions:
\begin{inparaenum}[(1)]
	\item verifying that bit substring of physical address is used to select cache sets and that it is possible to partition cache capacity based on set index;
	\item cracking hash function on Sandy Bridge 4 core processor and further the hash function to a simple function formula;
	\item cracking hash function on Sandy Bridge 6 core processor and presents the hash function in the form of 32 mapping tables. As different set indexes correspond to different mapping relationship, attentional attention are needed when designing the page coloring mechanism. However, because it's true that bit substring of physical address is used to select cache sets, although the hash function hasn't been reduced to a simple formula, it's still possible to implement cache partition based on cache set index.
\end{inparaenum}

This article is organized as follows: Section 2 defines the problem. Section 3 describes the procedure to crack cache hash function. Section 4 presents the observations based on which this article comes up with the assumption about the implementation of the hash function. Section 5 describes the assumptions about the implementations of the hash function. Section 6 presents the details of the cracking scheme. Section 7 presents the cracking results. Section 8 describes the details to verify the correctness of the cracking result. Section 9 describes the results of performance of cache partition implemented based on page coloring. Section 10 summarizes the contributions.
\begin{figure*}[!htbp]
\centering
\subfigure[Cache lines in one cache set are located on different positions on physical cache slice]{
		\centering
		\includegraphics[width=2.5in]{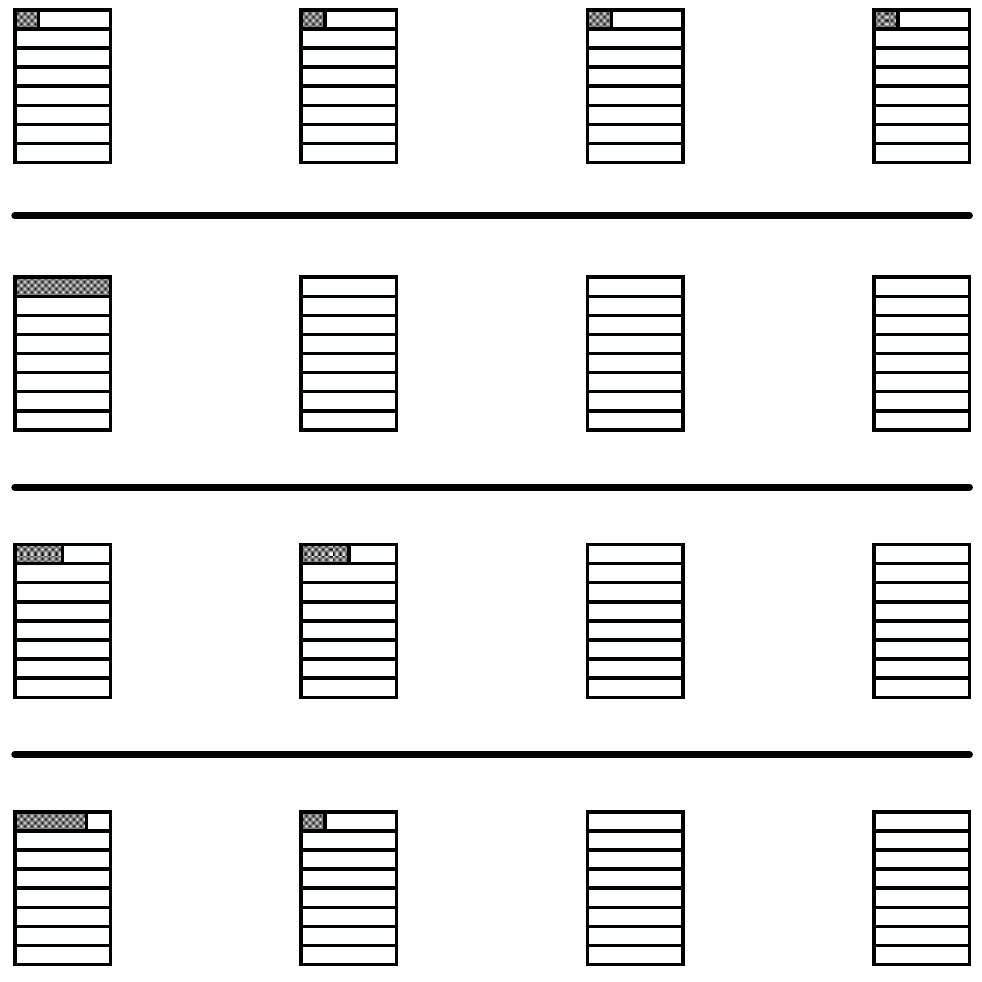}
	}
	~
	\subfigure[Consecutive cache sets are located on different positions on physical cache slice]{
		\centering
		\includegraphics[width=2.5in]{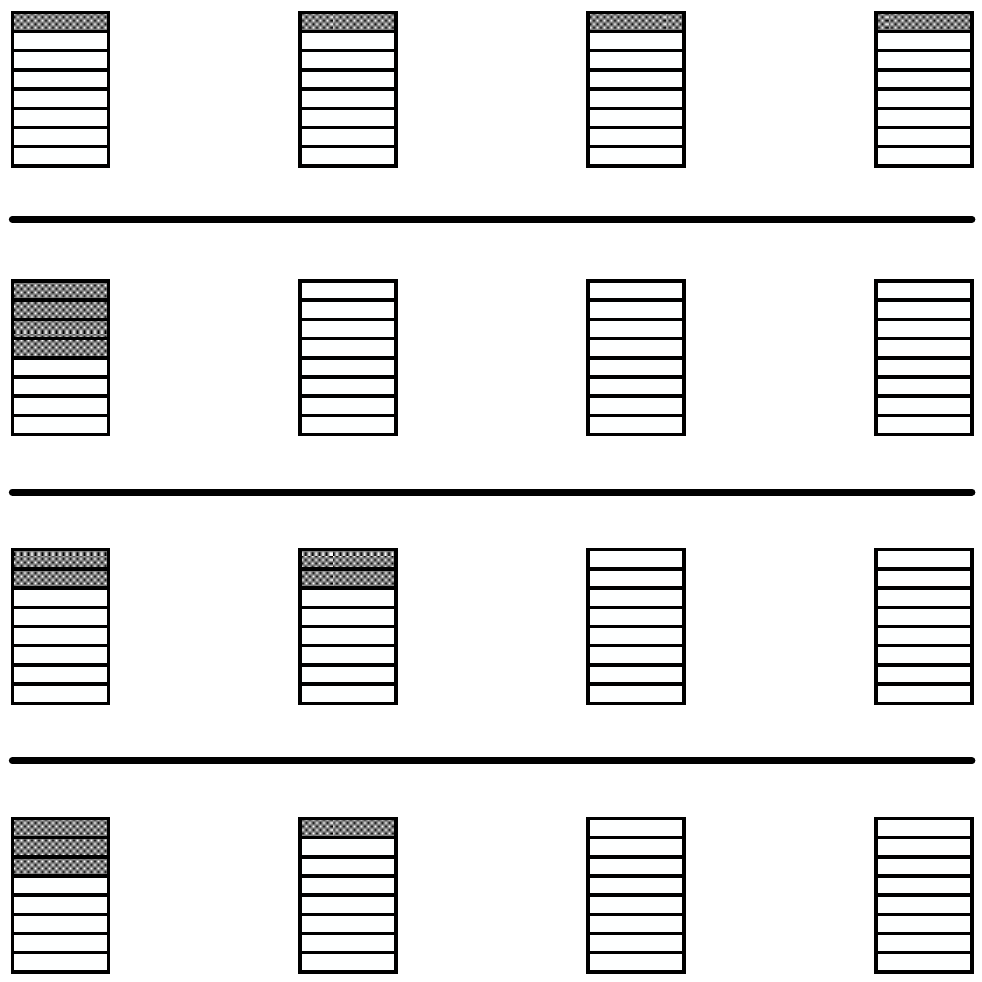}
	}
	\caption{Different mapping mechanism between cache set and physical cache.}
	\label{fig:PhysicalMappingScheme}
\end{figure*}
\section{Problem}
All the physical memory is divided into data blocks, and the size of data block is the same with the size of cache line.

A mapping function exists between data blocks and cache sets. For an unknown processor $P$, let $C_{cache}$ be the capacity of the cache, let $C_{memory}$ be the capacity of the memory installed, 
let $C_{cacheline}$ be the size of cache line, let $\Associativity$ be the associativity of the cache, and the total number of data block is 
\begin{equation}
	\NumberBlock = \frac{C_{memory}}{C_{cacheline}}
\end{equation}
the number of cache sets is 
\begin{equation}
	\NumberSet = \frac{C_{cache}}{a \times C_{cacheline}}
\end{equation}
On Sandy Bridge processor, LLC is also set-associative. What's different is that one hash function exists to distribute all data blocks across cache slices.
Let $location$ is the location on cache where a given $\MA$ is stored; as LLC is divided into slices, \emph{location} consists of two parts, $\SliceID$ and $\SetIndex$. $\SliceID$ is the LLC slice the data block is mapped to, and $\SetIndex$ is the index of the set the physical address is mapped to. 

Using these notations, this article defines the hash function on \SandyBridge processor as follows: given a physical address $\MA$, $\SliceID$ and $\SetIndex$, this article defines two functions $\MATOSliceID\left(\right)$ and $\MATOSetIndex\left(\right)$ to describe the relationship
\begin{equation}
	\SliceID = ma\_to\_slice\_id\left(\MA\right) 
\end{equation}
\begin{equation}
	\SetIndex = ma\_to\_set\_index\left(\MA\right)
\end{equation}
\section{Procedure}
\begin{figure}[!htbp]
\centering
\includegraphics[width=0.8\columnwidth]{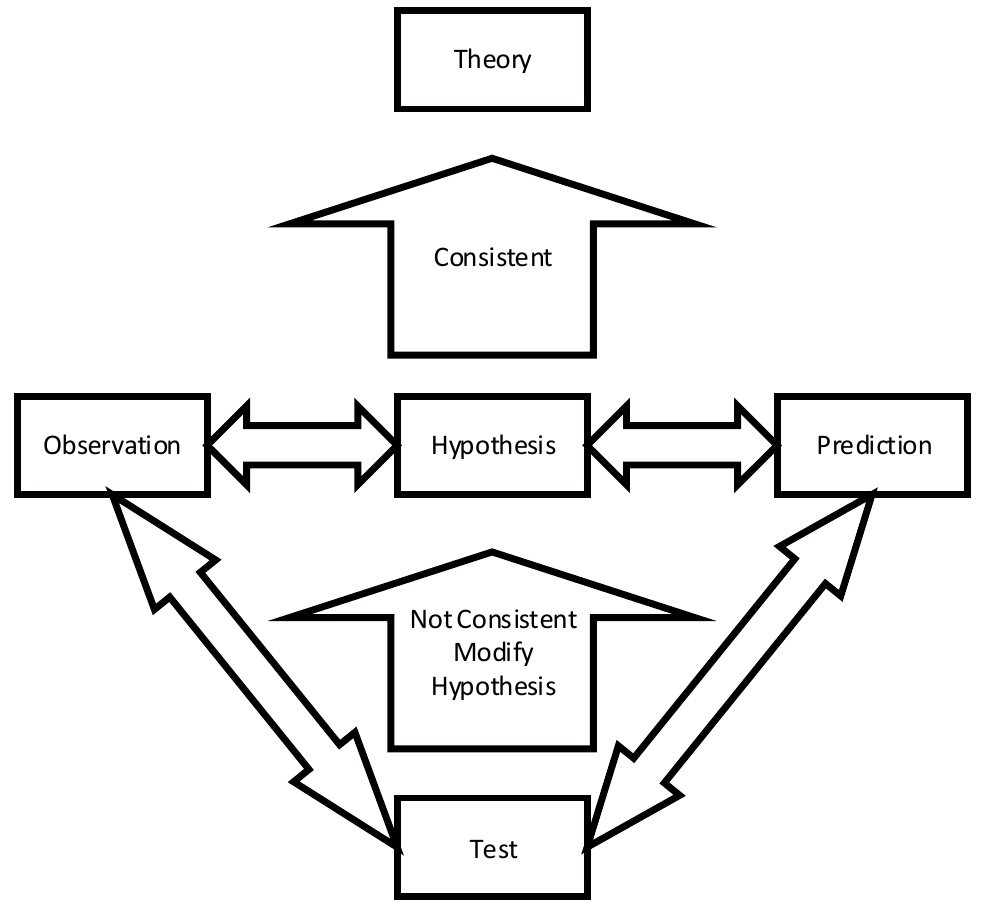}
\caption{The procedure to crack the hash function.}
\label{fig:TestProcedure}
\end{figure}

Figure~\ref{fig:TestProcedure} presents the procedure to crack the cache hash function on Intel Sandy Bridge processor. Based on some observations in prior work and some experiments, this article first presents a hypothesis. Then this article proves the hypothesis to be correct. 
\begin{figure*}[!htbp]
\centering
\subfigure[Stride varies from 64B to 64KB.]{
		\includegraphics[width=3in]{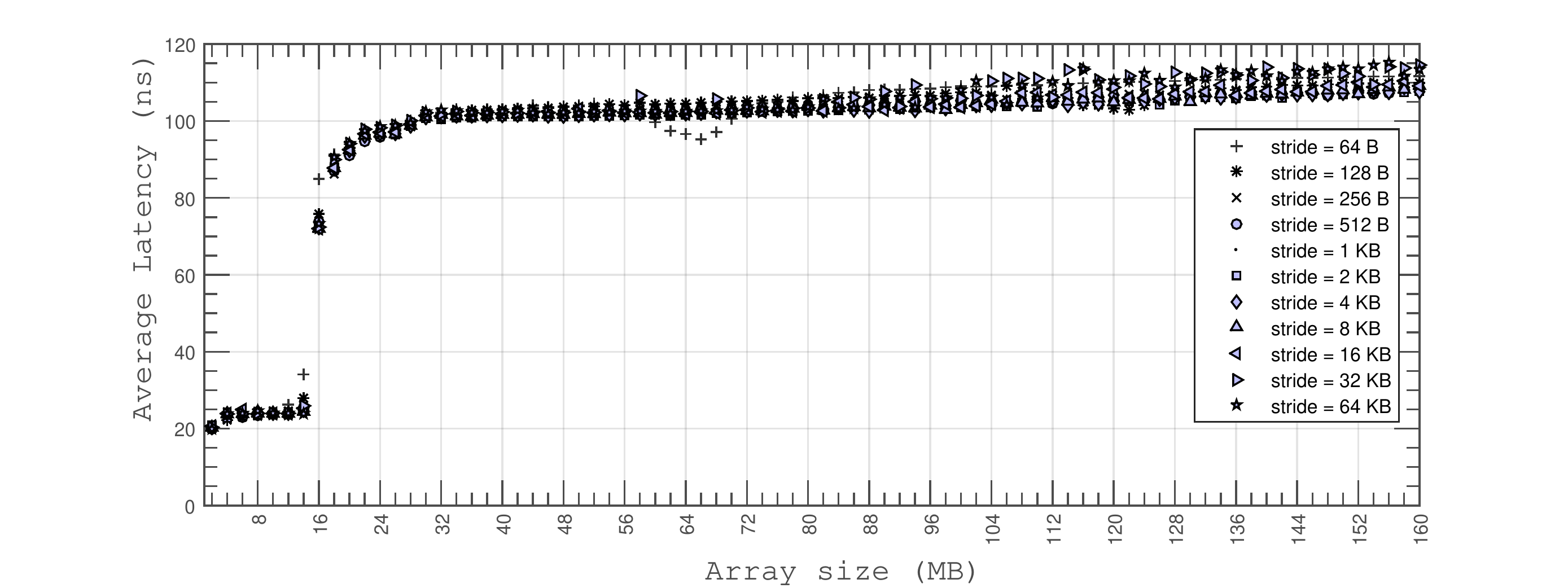}
		\label{fig:subfigure1_stride}
	}
	~
	\subfigure[Stride varies from 128KB to 1024KB]{
		\includegraphics[width=3in]{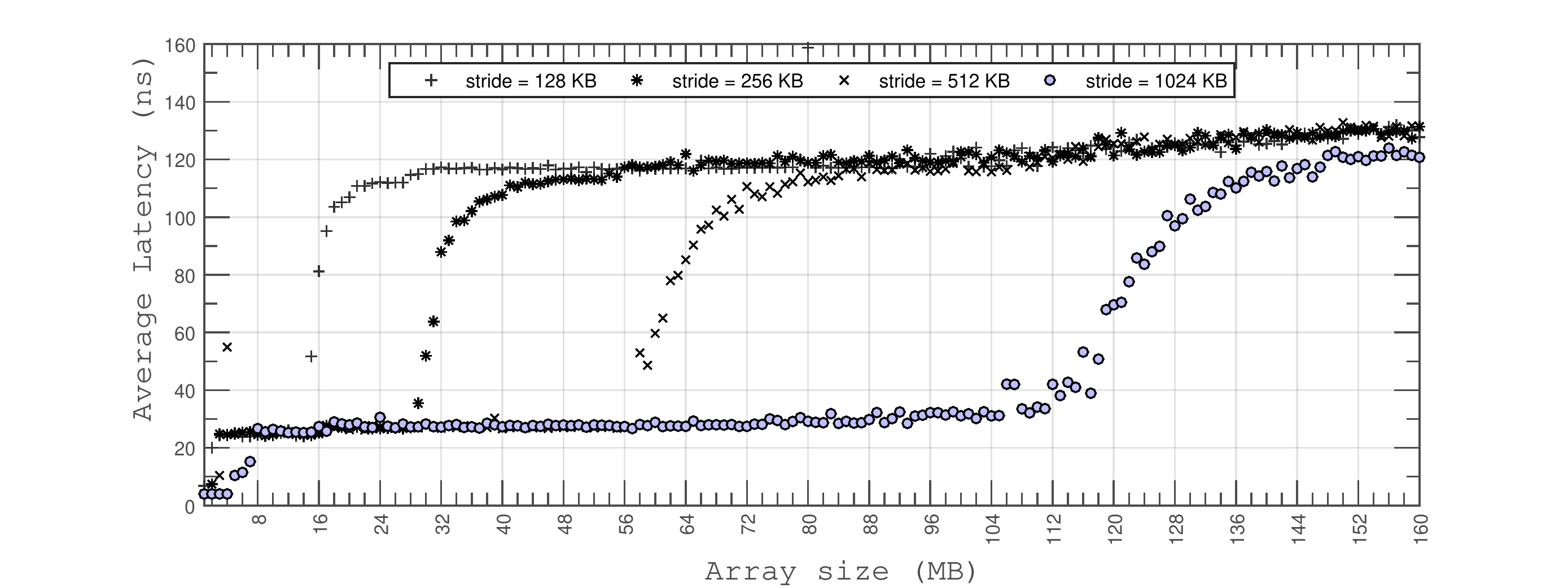}
		\label{fig:subfigure2_stride}
	}
        ~
        \subfigure[Stride varies from 2MB to 16MB.]{
		\includegraphics[width=3in]{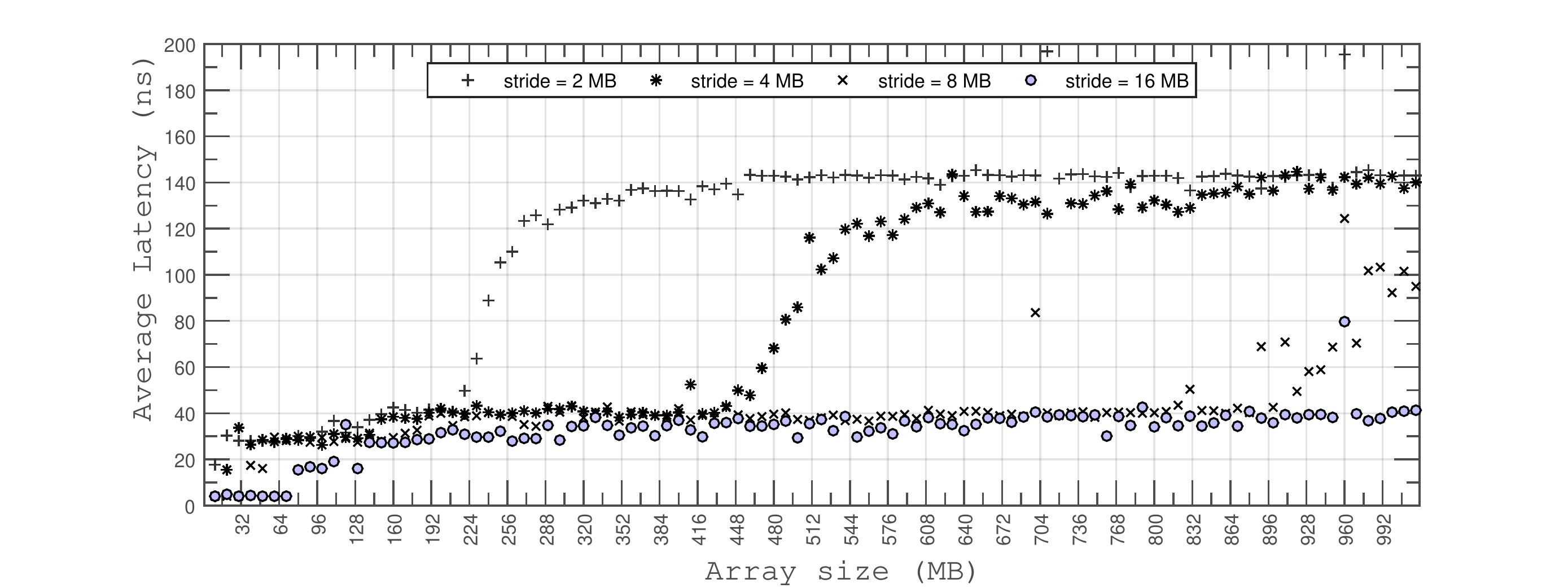}
		\label{fig:subfigure3_stride}	
	}
	~
	\subfigure[Stride varies from 32MB to 256MB.]{
		\includegraphics[width=3in]{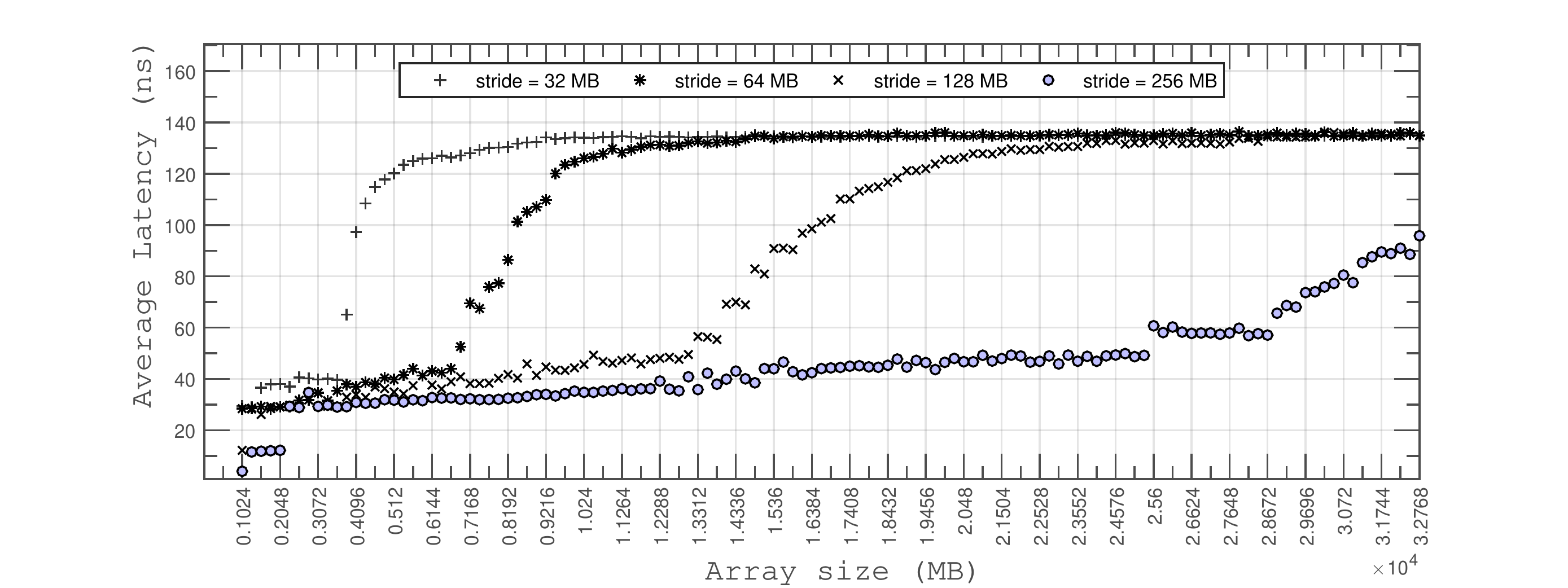}
		\label{fig:subfigure4_stride}
	}
	\caption{Perform data dependent access with different stride. For each stride, there is a point that when the average access latency begins to increase sharply.}
	\label{fig:SetIndexBits}
\end{figure*}

\begin{figure}[!htbp]
\centering
\includegraphics[width=0.8\columnwidth]{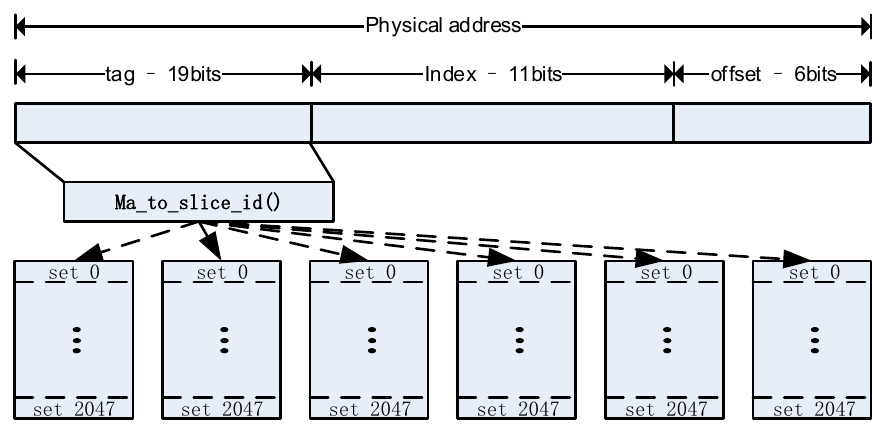}
\caption{Logical model about Intel Sandy Bridge LLC organization.}
\label{fig:LogicalModel}
\end{figure}

This article proposes a logical model to describe the problem clearly. As presented in~\cite{Multi-core-cache_rajeev}, physical cache has complex organization. The method discussed in this article isn't able to distinguish between the following situations as presented in figure~\ref{fig:PhysicalMappingScheme}. In order to describe the problem clearly, this article proposes a logical model of cache organization, as presented in figure~\ref{fig:LogicalModel}. In this model, each slice is divided into different cache sets, the number of cache sets on each cache slice can be decided based on the capacity of each cache slice, the size of cache line and associativity of cache set. Each cache slice consists of the same number of cache sets. One exact cache set is selected by specifying \SliceID and \SetIndex. Besides, without specifying \SliceID, the cache sets with the same \SetIndex  on each cache slice will be selected. 
\section{Observation}
\subsection{Main memory and LLC access latency}
When the data accessed can't be held in cache, cache miss will cause average access latency to increase. 
\subsection{Substring of physical address serves as set index}
As described in section methodology, the test program is able to access data blocks in specified cache sets. This article first accesses the physical memory with different stride, and the size of physical memory and the stride is recorded. Besides, when the number of accessed data blocks exceeds the number of the selected cache sets, serious conflict will cause average access latency to increase sharply.
\begin{table}[!htbp]
	\centering
	\caption{The number of cache lines that reside in cache when accessing using different stride and array size.}
	\begin{tabular}{cccc}
		\toprule
		stride  &number    & stride & number \\ \otoprule
		64B      & $15 \times 2^{14}$  &   32KB      &   $15 \times 2^{5}$     \\ \midrule
		128B     & $15 \times 2^{13}$  &   64KB      &   $15 \times 2^{4}$     \\ \midrule
		256B     & $15 \times 2^{12}$  &  128KB      &   $15 \times 2^{3}$     \\ \midrule
		512B     & $15 \times 2^{11}$  &  256KB      &   $15 \times 2^{3}$     \\ \midrule
		 1KB     & $15 \times 2^{10}$  &  512KB      &   $15 \times 2^{3}$     \\ \midrule
		 2KB     & $15 \times 2^{9}$   &    1MB      &   $15 \times 2^{3}$     \\ \midrule
		 4KB     & $15 \times 2^{8}$   &    2MB      &   $15 \times 2^{3}$     \\ \midrule
		 8KB     & $15 \times 2^{7}$   &    4MB      &   $15 \times 2^{3}$     \\ \midrule
		16KB     & $15 \times 2^{6}$   &    8MB      &   $15 \times 2^{3}$     \\ \bottomrule
		\end{tabular}
		\label{tab:SetIndexBitsSingularity}
\end{table}

As presented in figure~\ref{fig:SetIndexBits}, average access latency increases sharply at one point. This point represents the configuration of the test, including array size and stride. With each configuration, the number of data blocks accessed is decided by array size and stride. As shown in figure~\ref{tab:SetIndexBitsSingularity}, when the stride is larger than 32KB, the number of cache lines LLC can hold equals to 120. Chances are that some bits in physical address serves as set index during cache access. When the stride is large enough, the set index will remain unchanged for the data blocks accessed. Considering the fact that the associativity of the \SandyBridge processor in this test is 20, and has $6 \times 20 = 120$ cache sets. 
\subsection{The hash function meets some properties}
As presented in~\cite{seznec1993case}, the set-associative organization cache should meet the following properties to provide better performance.
\begin{inparaenum}[(1)]
\item Equitability;
\item Local dispersion;
\item Simple hardware implementation;
\end{inparaenum}
\begin{table*}[!htbp]
\centering
\caption{Processor parameters.}
\begin{tabular}{lrr} 
	\toprule
	CPU Type	&Intel\textregistered Xeon\textregistered Processor E5-2640&Intel\textregistered Xeon\textregistered Processor E5-2603  \\ \otoprule
	Intel CPU core	&6 cores@2.5G			& 4 cores@2.5G										\\ 
	L1 I-Cache 	& 32kB/core, 8-way, 64B line   	& 32kB/core, 8-way, 64B line								\\ 
	L1 I-Cache 	& 32kB/core, 8-way, 64B line   	& 32kB/core, 8-way, 64B line								\\ 
	L2 Cache   	& 256kB/core, 8-way, 64B line  	& 256kB/core, 8-way, 64B line								\\ 
	L3 Cache   	&15360kB(shared, 6 slices); 20-way, 64B line&10240kB(shared, 4 slices); 20-way, 64B line				\\ 
	Memory Capacity & 64GB & 16GB \\ \bottomrule
\end{tabular}
\label{table:ServerConfiguration}
\end{table*}
\section{Assumption}
\begin{figure}[!htbp]
\centering
\includegraphics[width=0.8\columnwidth]{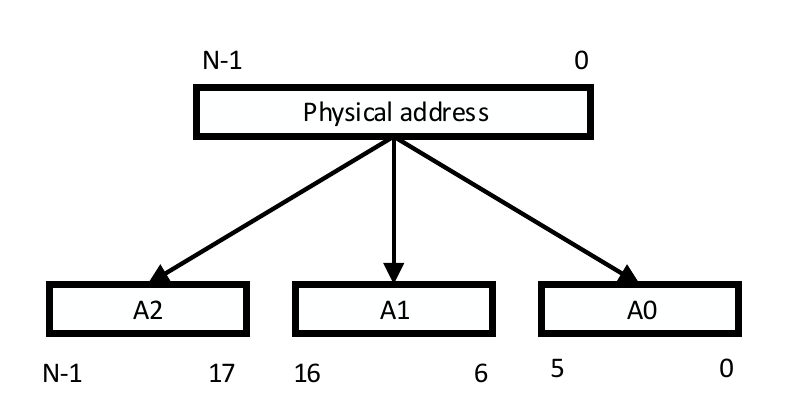}
\caption{Split physical address into different fields.}
\label{fig:SplitPhysicalAddress}
\end{figure}

Let $\NUMSLICE$ be the total number of slices and $\NUMSET$ be the number of sets on each slice, 
$\left(a_{\PhysicalAddressWidth-1},\dots,a_0\right)$ be the binary representation of the physical address. 
As presented in figure~\ref{fig:SplitPhysicalAddress}, this article splits the binary representation of an address $\MA$ into bit substrings $\left(A_2, A_1, A_0\right)$, $A_0$ is a $c$ bit string: the displacement in the line. 
$A_1$ is a \WIDTHSetIndex{} bit string and $A_2$ is the string of the most significant bits. 
Based on the work presented in~\cite{sandybridgehash}, this article makes the following assumptions:
\begin{enumerate}
\item the value of $A_0$ is used as block address;
\item the value of $A_1$ is used as set index on a specified cache slice;
\item $A_2$ is used to decide the $\SliceID$ of a given $\MA$;
\end{enumerate}
\section{Methodology}
This section discusses the following three questions: 
\begin{inparaenum}[(1)]
\item the criteria which is used to classify all the physical addresses;
\item the mechanism to ensure the correctness of the criteria;
\item the method to get the information needed to perform the classification of data blocks. 
\end{inparaenum}
\subsection{Platform}
Table~\ref{table:ServerConfiguration} presents the parameters of the processors used in this article.
\subsection{Classifying criteria}
\begin{figure*}
\centering
\includegraphics[width=5in]{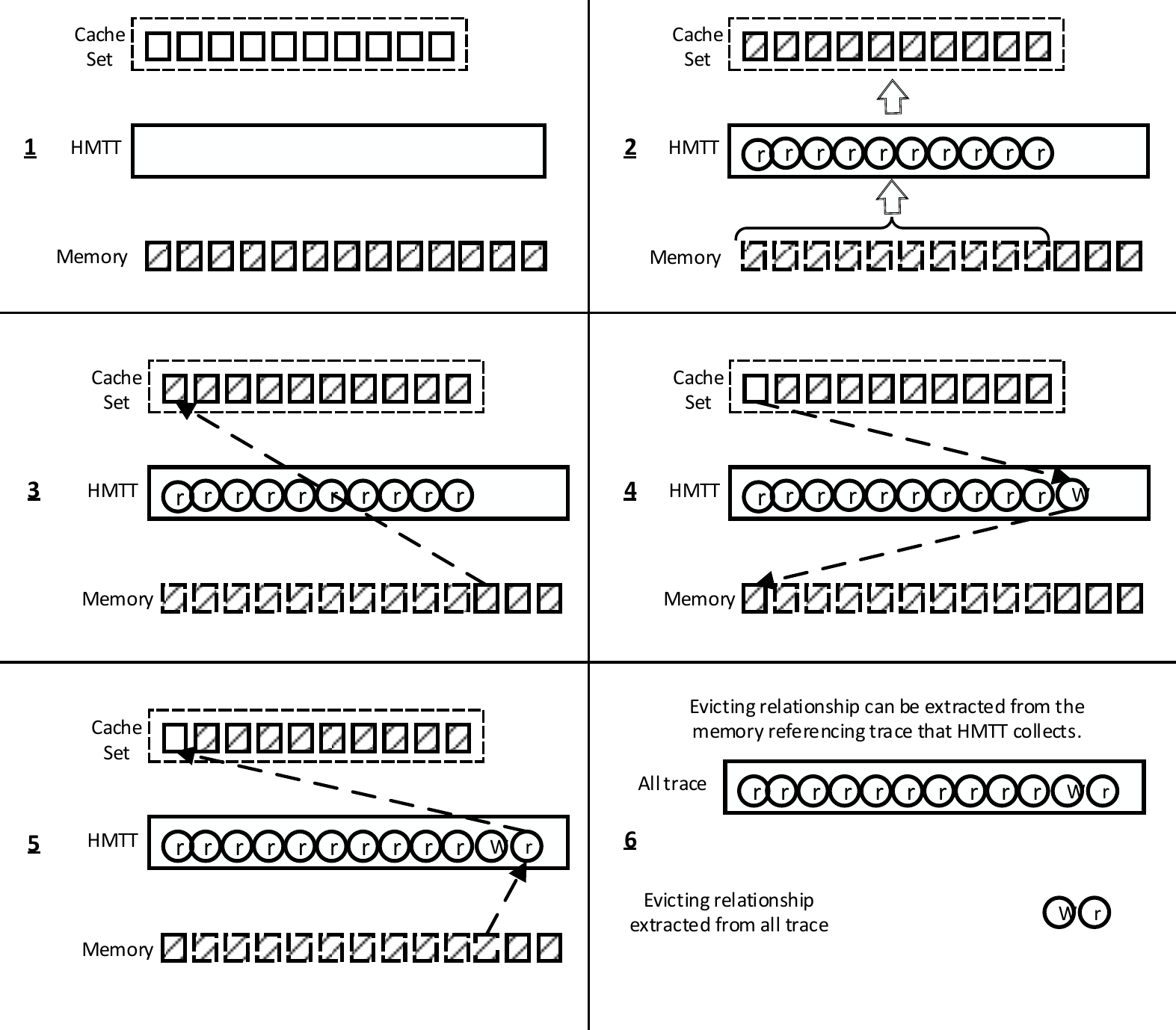}   
\caption[1]{Extract evicting relationship between data block from memory reference trace that HMTT collects:
\begin{inparaenum}[(1)] 
\item The array is initialized in a data dependent manner; 
\item \Associativity data blocks are read into cache; 
\item the next data block is going to be read; 
\item one cache line is evicted before the next data block can be read into cache; 
\item next data block is read into cache. 
\end{inparaenum}
}
\label{fig:New09CacheReplacementHmttTrace}
\end{figure*}

Evicting relationship exists between \WordDataBlock s that are mapped to the same cache set. When the number of accessed data blocks exceeds the associativity of cache set, cache conflict occur, these \WordDataBlock will begin to evict each other, as presented in figure~\ref{fig:New09CacheReplacementHmttTrace}. This article uses this evicting relationship to classify \WordDataBlock into different groups. 
\subsection{Getting the evicting relationship between \WordDataBlock}
\subsubsection{Accessing data in desired cache set}
In order to get the evicting relationship, the testing program should be able to fill data into specified cache set. We boot the operating system with 2GB memory. In this way, the OS can only use the lower 2GB memory. This article further implement a driver to map the other physical memory ranging from 2GB to 3GB into kernel space. In this way, the physical address accessed can be calculated by subtracting based address from the linear address of the operating system. 
\begin{table}[!htbp]
	\caption{Use combinations of bits to generate addresses}
	\centering
	\subtable[bit $a_2, a_1, a_0$]{
		\begin{tabular}{cc}
		\toprule
		Bit value&Value \\ \otoprule
		000&0x0 \\ \midrule
		001&0x1 \\ \midrule
		010&0x2 \\ \midrule
		011&0x3 \\ \midrule
		100&0x4 \\ \midrule
		101&0x5 \\ \midrule
		110&0x6 \\ \midrule
		111&0x7 \\ \bottomrule
		\end{tabular}
	}
	\qquad
	\subtable[bit $a_{18}, a_{17}, a_{16}$]{
		\begin{tabular}{cc}
		\toprule
		Bit value&Value \\ \otoprule
		000&0x00000 \\ \midrule
		001&0x10000 \\ \midrule
		010&0x20000 \\ \midrule
		011&0x30000 \\ \midrule
		100&0x80000 \\ \midrule
		101&0x90000 \\ \midrule
		110&0xa0000 \\ \midrule
		111&0xb0000 \\ \bottomrule
		\end{tabular}
	}
	\label{table:3bitscombination}
\end{table}
\subsubsection{Test sequence generation} 
As presented in table~\ref{table:3bitscombination}, the physical address is generated by bit combination. In this way, this article verifies the effect of every bit on the result of hash function. 
\begin{figure}[!htp]
\centering
\lstset { %
	language=C,
	backgroundcolor=\color{black!5}, 
	caption={The main opration of test program without interval between consecutive acesses.},
	captionpos=b,
	frame=single,
	label={snippetwithoutmakedirty},
	basicstyle=\footnotesize,
}
\begin{lstlisting}
    while (j < iteration) {
        addr = *(TYPE*)(addr);
        j++;
    }
\end{lstlisting}
\end{figure}
\begin{figure}[!htp]
	\centering
\lstset { %
	language=C,
	backgroundcolor=\color{black!5}, 
	caption={The main opration of test program with interval between consecutive acesses.},
	captionpos=b,
	frame=single,
	label={snippetwithmakedirty},
	basicstyle=\footnotesize,
}
\begin{lstlisting}
    while (j < iteration) {
        addr = *(TYPE*)(addr);
        //modify the cache line
        *(TYPE*)(addr+16) = 1; 
        while (k++ < 1000);
        j++;
        k=0;
    }
\end{lstlisting}
\end{figure}
\subsubsection{Array initialization}
\begin{figure}[!htbp]
\centering
\includegraphics[width=0.8\columnwidth]{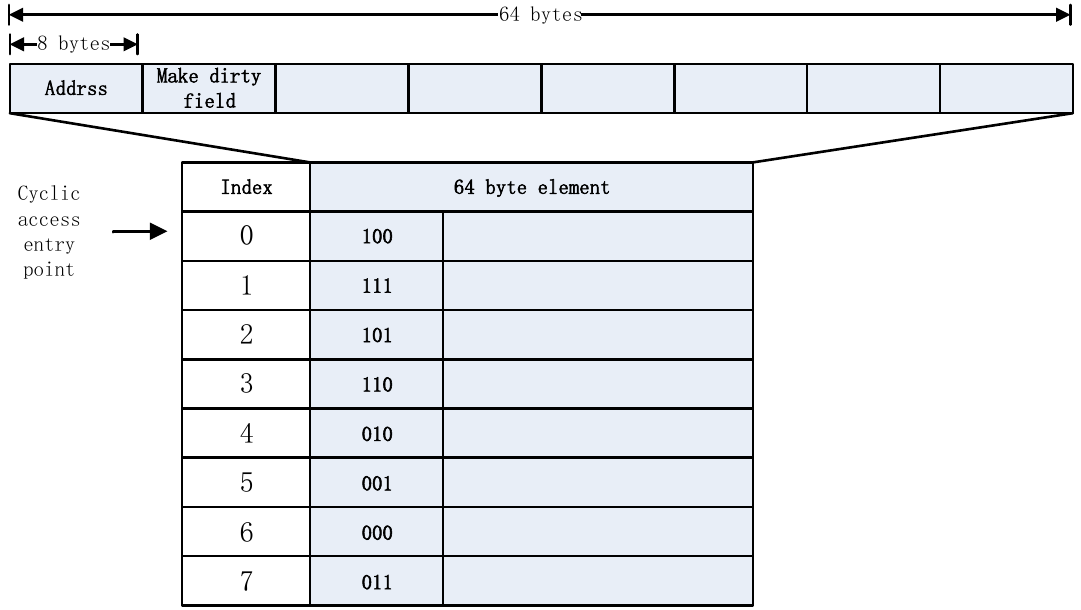}
\caption{The test array is initialized using shuffled addresses in a data-dependent manner.}
\label{fig:cyclicaccess}
\end{figure}

The testing program firstly allocates an array. Then the testing program initializes the test array in a data-dependent manner, which means that the data stored at the current read address is the address of the next read command. As shown in Figure~\ref{fig:cyclicaccess}, the main operation of the test program is to read data and use the data as the address of the following read operation. 
\subsubsection{Ensure the correctness of evicting relationship}
As depicted in code snippet~\ref{snippetwithmakedirty}, idle loop is inserted between adjacent memory accesses to make sure that only two physical addresses with eviction relationship exists between each other are temporally adjacent.
\subsubsection{Collecting memory reference trace}
HMTT is a hybrid hardware/software memory trace monitoring system. This tool collects all the memory reference trace, which memory controller issues to memory modules. Although the tool can be programmed to collect various information~\cite{chen2014cmd}, the information needed in this article includes physical address, time interval between different consecutive physical addresses and read/write bit of the physical address, as presented in table~\ref{table:SampleMemoryReferenceTrace}. The whole process is presented in figure~\ref{fig:New09CacheReplacementHmttTrace}. When the cache set is full and testing program reads another data block into cache, one of the data blocks already read into cache needed to be evicted to make room for the newly read data block. The two operations are collected by HMTT and save as two memory reference trace. The trace consists of information about the address of the operation and whether the operation is read or write. 
\subsubsection{Classify all the physical address into different groups}
\begin{figure*}[!htbp]
\centering
\subfigure[Breadth first search.]{
		\centering
		\includegraphics[width=2.5in]{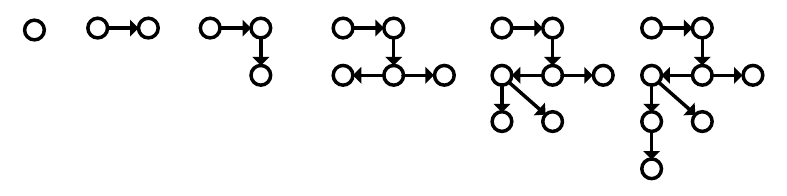}
		\label{fig:ConnectedGraphBreadthFirst}
	}
	~
	\subfigure[Represent data block with a node, and Represent evicting relationship with an edge.]{
		\centering
		\includegraphics[width=2.5in]{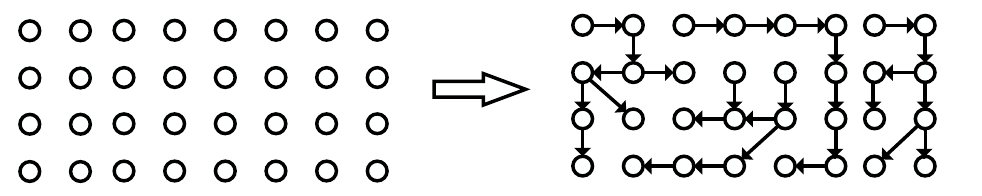}
		\label{fig:ConnectedGraphNodeEdge}
	}
	\subfigure[The data blocks are classified into different groups.]{
		\centering
		\includegraphics[width=2.5in]{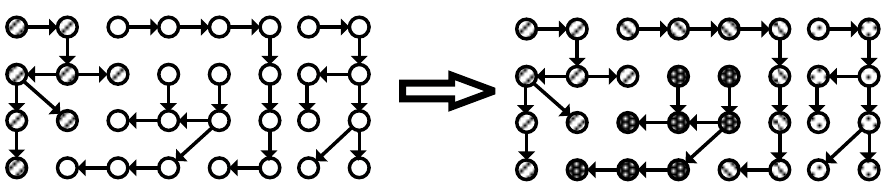}
		\label{fig:ConnectedGraphResult}
	}
	\caption{This article uses connected subgraph method to solve the problem of classifying \WordDataBlock.}
\end{figure*}

As presented in figure~\ref{fig:ConnectedGraphNodeEdge}, if evicting relationship exists between physical address A and B, and also exists between physical address B and C, then it is concluded that evicting relationship also exists between physical address B and C. This article further classifies all the physical addresses into different groups based on connected subgraph related method. 
\clearpage
\begin{table}[!htbp]
\caption{Sample memory reference trace.}
\begin{center}
\begin{tabular}{cccc} 
\toprule
Seq	& Read or Write	&	Physical Address&	Interval \\ \otoprule
1	&	read	&	bfd60000	&	15	 \\ \midrule
2	&	write	&	be1a0000	&	1094	 \\ \midrule
3	&	read	&	bfd80000	&	15	 \\ \midrule
4	&	write	&	be4a0000	&	608	 \\ \midrule
5	&	read	&	bfda0000	&	9	 \\ \midrule
6	&	write	&	be500000	&	1206	 \\ \midrule
7	&	read	&	bfdc0000	&	20	 \\ \midrule
8	&	write	&	bef40000	&	1090	 \\ 
\bottomrule
\end{tabular}
\end{center}
\label{table:SampleMemoryReferenceTrace}
\end{table}
\begin{table*}
	\begin{subtable}{}
		\begin{align*}
			bit\_a_0\ =\ &get\_bit\left(A2,\ 0\right)\ \oplus\ get\_bit\left(A2,\ 1\right)\ \oplus\ get\_bit\left(A2,\ 2\right)\ \oplus\ get\_bit\left(A2,\ 3\right)\ \oplus\ get\_bit\left(A2,\ 4\right)\ \oplus\  \\ 
		 	&get\_bit\left(A2,\ 5\right)\ \oplus\ get\_bit\left(A2,\ 7\right)\ \oplus\ get\_bit\left(A2,\ 9\right)\ \oplus\ get\_bit\left(A2,\ 10\right)\ \oplus\ \left(get\_bit\left(A2,\ 12\right)\ \right. \&\  \\ 
		  	&\left.get\_bit\left(A2,\ 14\right)\right)\ \oplus\ \left(\ \left(\sim get\_bit\left(A2,\ 14\right)\right)\ \&\ get\_bit\left(A2,\ 13\right)\ \right)\ 
		\end{align*}
	\end{subtable}
	~
\begin{subtable}{}
		\begin{align*}
			bit\_a_1\ =\ &get\_bit\left(A2,\ 0\right)\ \oplus\ get\_bit\left(A2,\ 2\right)\ \oplus\ get\_bit\left(A2,\ 4\right)\ \oplus\ get\_bit\left(A2,\ 6\right)\ \oplus\ get\_bit\left(A2,\ 8\right)\ \oplus\ \\
			             &get\_bit\left(A2,\ 10\right)\ \oplus\ get\_bit\left(A2,\ 11\right)\ \oplus\ get\_bit\left(A2,\ 13\right)\ \oplus\ \\ 
			             &\left(get\_bit\left(A2,\ 14\right)\ \&\ get\_bit\left(A2,\ 13\right)\ \&\ get\_bit\left(A2,\ 12\right)\right)\ 
		\end{align*}
	\end{subtable}
	\caption{Two intermediate value used to reduce Sandy Bridge 4 core mapping table.}
	\label{fig:TwoIntermediateValue}
\end{table*}
\section{Results}
This article presents the mapping relationship in the form of mapping table. It's true with both \SandyBridge 4 core and 6 core processor that bit string $A1$ in physical address selects cache set directly.
For both processor, there are $\NumberSetEachSlice$ cache sets per cache slice. The number of data blocks that are mapped to cache set with the same set index is  
\begin{equation}
	\frac{\MemoryCapacity}{C_{\WordDataBlock} \times \NumberSetEachSlice} 
\end{equation}
Let $\MemoryCapacity$ be the installed memory, let $\DataBlockCapacity$ the size of data block, and let $\NumberBlock$ be total number of data blocks, due to the fact that, as presented in figure~\ref{fig:SplitInto4Simplication}, substring A1 in physical address is used to select cache set, for those data blocks those share the same set index, they will be mapped to cache sets on all cache slice sharing the same cache set index, the total number of cache slice is $\NUMSLICE$. For those blocks sharing the same set index, this article uses a mapping table to describe the relationship.
\subsection{4 core processor}
On \SandyBridge 4 core processor, the installed memory is 16GB, so the total data block is 
\begin{equation}
	\NumberBlock = \frac{\MemoryCapacity}{\DataBlockCapacity} = \frac{16 GB}{64 B} = 2^{28}
\end{equation}
The number of cache sets on each cache slice is 2048, the number of data block that share the same cache set index is 
\begin{equation}
	\frac{\NumberBlock}{\NUMSET} = \frac{2^{28}}{2^{11}} = 2^{17}
\end{equation}
%
%
%
\begin{table}[!h]
\centering
\caption{4 core cache hash mapping table}
\label{tab:4CoreHashTable}
\begin{tabular}{cccc} 
	\toprule
Slice 0 & Slice 1 & Slice 2 & Slice 3 \\ \otoprule
4000 & 4001 & 4002 & 4003  \\ \midrule
4007 & 4006 & 4005 & 4004  \\ \midrule
4009 & 4008 & 400b & 400a  \\ \midrule
400e & 400f & 400c & 400d  \\ \midrule
4013 & 4012 & 4011 & 4010  \\ \midrule
4014 & 4015 & 4016 & 4017  \\ \midrule
401a & 401b & 4018 & 4019  \\ \midrule
401d & 401c & 401f & 401e  \\ \midrule
4021 & 4020 & 4023 & 4022  \\ \midrule
4026 & 4027 & 4024 & 4025  \\ \midrule
4028 & 4029 & 402a & 402b  \\ \midrule
402f & 402e & 402d & 402c  \\ \midrule
4032 & 4033 & 4030 & 4031  \\ \midrule
4035 & 4034 & 4037 & 4036  \\ \midrule
403b & 403a & 4039 & 4038  \\ \midrule
403c & 403d & 403e & 403f  \\ 
\bottomrule
\end{tabular}
\end{table}
These data blocks will be mapped to these selected cache sets. As presented in table~\ref{tab:4CoreHashTable}, as all the blocks share the same set index, this article only presents the A2 string of each block address here. 
Each set index corresponds to a mapping table. This article finds that on Intel 4 core processor, the mapping tables corresponds to different set index are the same.

The installed physical memory is 16GB, so we have the mapping table of 34 bit width physical address. Because bit string A1 in physical address selects cache set directly, there are 2048 cache sets on each cache slice. As a result, there should have been 2048 mapping table to describe the relationship. However, this article finds that the mapping table is the same for all set index. 

\begin{figure*}[!htbp]
\centering
\subfigure[Figure A]{
		\centering
		\includegraphics[width=2.5in]{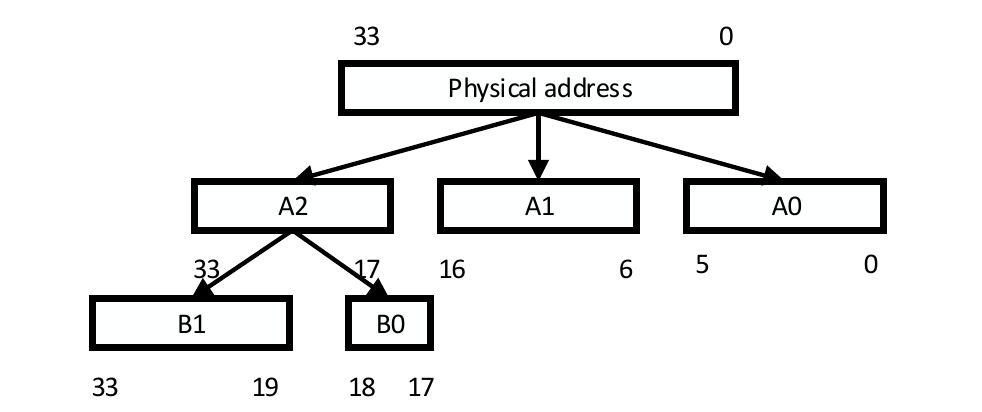}
		\label{fig:SplitInto4Simplication}
	}
	~
	\subfigure[Figure B]{
		\centering
		\includegraphics[width=2.5in]{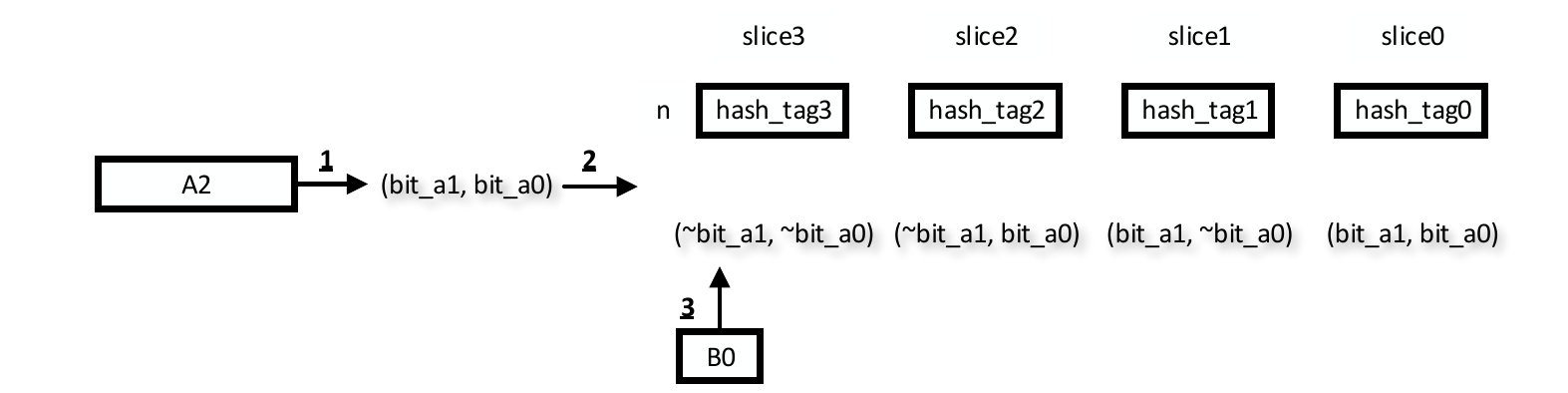}
		\label{fig:DecideBlockLocation}
	}
	\caption{Intel Sandy Bridge 4 core processor hash fucntion cracking result.}
\end{figure*}

Reduction of \SandyBridge processor mapping table. The mapping table can be reduced to a simple formula. 
As presented in figure~\ref{fig:DecideBlockLocation}, two intermediate value~\ref{fig:TwoIntermediateValue}, $bit\_a_0$, $bit\_a_1$, these four value is related to four different cache slices. The value of $bit_{16}bit_{15}$ is used to select one set from from four cache sets selected by set index of the data block. 
\subsection{6 core processor}
On \SandyBridge 6 core processor, the installed memory is 64GB, so the total data block is 
\begin{equation}
	\NumberBlock = \frac{\MemoryCapacity}{C_{\WordDataBlock}} = \frac{64 GB}{64 B} = 2^{30}
\end{equation}
Because there are 2048 cache sets on each slice, the number of data block sharing the same cache set index is 
\begin{equation}
	\frac{\NumberBlock}{\NUMSET} = \frac{2^{30}}{2^{11}} = 2^{19} 
\end{equation}

\begin{table*}
\caption{Intel Sandy Bridge 6 core processor hash function cracking result.}
\centering
\subtable[6 core cache hash mapping table, set index 1]{
	\begin{tabular}{cccccc} 
		\toprule
	Slice 0       & \multicolumn{1}{c}{Slice 1} & \multicolumn{1}{c}{Slice 2} & \multicolumn{1}{c}{Slice 3} & \multicolumn{1}{c}{Slice 4} & \multicolumn{1}{c}{Slice 5} \\ \otoprule	
	4000 & 4001  & 4002  & 4003  & 4007  & 400e \\ \midrule
	400d & 4006  & 4005  & 4004  & 400a  & 400f \\ \midrule
	4017 & 400c  & 4008  & 4009  & 400b  & 4014 \\ \midrule
	401a & 4016  & 4012  & 4013  & 4010  & 4015 \\ \midrule
	401d & 401b  & 401f  & 401e  & 4011  & 4018 \\ \midrule
	4023 & 401c  & 4026  & 4027  & 4024  & 4019 \\ \midrule
	4029 & 4022  & 402b  & 402a  & 4025  & 4020 \\ \midrule
	402e & 4028  & 402c  & 4030  & 4032  & 4021 \\ \midrule
	4034 & 402f  & 4031  & 4037  & 4033  & 402d \\ \midrule
	4039 & 4035  & 4036  & 403d  & 403e  & 403a \\ \midrule
	4041 & 4038  & 403c  & 4042  & 403f  & 403b \\ \midrule
	4046 & 4040  & 4043  & 4045  & 404c  & 4044 \\ \midrule
	404b & 4047  & 404e  & 404f  & 404d  & 4048 \\ \midrule
	4051 & 404a  & 4054  & 4055  & 4056  & 4049 \\ \midrule
	405c & 4050  & 4059  & 4058  & 4057  & 4052 \\ \midrule
	4065 & 405d  & 405e  & 405f  & 405a  & 4053 \\ \midrule
	4068 & 4064  & 4060  & 4061  & 405b  & 4066 \\ \midrule
	406f & 4069  & 406a  & 406b  & 4062  & 4067 \\ \midrule
	4072 & 4073  & 406d  & 406c  & 4063  & 4070 \\ \midrule
	4075 & 4074  & 4077  & 4076  & 406e  & 4071 \\ \midrule
	407f & 407e  & 407a  & 407b  & 4078  & 407c \\ \midrule
	NULL & NULL  & NULL  & NULL  & 4079  & 407d \\ 
	\bottomrule
	\end{tabular}
}
\qquad
\subtable[6 core cache hash mapping table, set index 2]{
	\begin{tabular}{cccccc} 
		\toprule
	Slice 0       & \multicolumn{1}{c}{Slice 1} & \multicolumn{1}{c}{Slice 2} & \multicolumn{1}{c}{Slice 3} & \multicolumn{1}{c}{Slice 4} & \multicolumn{1}{c}{Slice 5} \\ \otoprule
	4000 & 4002 & 4003 & 4004 & 4006 & 4007  \\ \midrule
	4001 & 4008 & 4009 & 4005 & 400b & 400a  \\ \midrule
	400c & 400f & 400e & 4012 & 4011 & 4010  \\ \midrule
	400d & 4015 & 4013 & 401e & 4016 & 4017  \\ \midrule
	401a & 4018 & 4014 & 401f & 401c & 401d  \\ \midrule
	401b & 4021 & 4019 & 4026 & 4022 & 4023  \\ \midrule
	402e & 402c & 4020 & 4027 & 4025 & 4024  \\ \midrule
	402f & 4036 & 402d & 402a & 4028 & 4029  \\ \midrule
	4034 & 403b & 4037 & 402b & 4032 & 4033  \\ \midrule
	4035 & 403c & 403a & 4030 & 403f & 4039  \\ \midrule
	4038 & 4044 & 403d & 4031 & 4040 & 403e  \\ \midrule
	4046 & 4049 & 4045 & 4042 & 404a & 4041  \\ \midrule
	4047 & 4053 & 4048 & 4043 & 404d & 404b  \\ \midrule
	4051 & 4054 & 4052 & 404e & 4050 & 404c  \\ \midrule
	405c & 405e & 4055 & 404f & 4057 & 4056  \\ \midrule
	405d & 4060 & 405f & 4058 & 405a & 405b  \\ \midrule
	4064 & 4067 & 4061 & 4059 & 4063 & 4062  \\ \midrule
	4065 & 406a & 4066 & 406c & 406e & 406f  \\ \midrule
	4068 & 4070 & 406b & 406d & 4074 & 4075  \\ \midrule
	4069 & 407a & 4071 & 4076 & 4079 & 4078  \\ \midrule
	4072 & 407d & 407c & 4077 & 407e & 407f  \\ \midrule
	4073 & NULL & NULL & NULL & 407b & 407d  \\ 
	\bottomrule
	\end{tabular}
}
\end{table*}
\begin{table*}[!htbp]
\centering
\caption{Mapping table corresponding to set index ranging from 0 to 127 shares 32 different mapping table. The situation is the same for set index ranging from 128 to 2047.}
\begin{tabular}{cccccccccc} 
	\toprule
	Mapping table index & Index & Index &  Index & Index & Mapping table index & Index & Index &  Index & Index    \\ \otoprule
 1   &   0   &  2    &   65  &   67 & 17  &   32  &  34   &   97  &   99 \\ \midrule    
 2   &   1   &  3    &   64  &   66 & 18  &   33  &  35   &   96  &   98 \\ \midrule
 3   &   4   &  6    &   69  &   71 & 19  &   36  &  38   &   101 &  103 \\ \midrule
 4   &   5   &  7    &   68  &   70 & 20  &   37  &  39   &   100 &  102 \\ \midrule
 5   &   8   &  10   &   73  &   75 & 21  &   40  &  42   &   105 &  107 \\ \midrule
 6   &   9   &  11   &   72  &   74 & 22  &   41  &  43   &   104 &  106 \\ \midrule
 7   &   12  &  14   &   77  &   79 & 23  &   44  &  46   &   109 &  111 \\ \midrule
 8   &   13  &  15   &   76  &   78 & 24  &   45  &  47   &   108 &  110 \\ \midrule
 9   &   16  &  18   &   81  &   83 & 25  &   48  &  50   &   113 &  115 \\ \midrule
 10  &   17  &  19   &   80  &   82 & 26  &   49  &  51   &   112 &  114 \\ \midrule
 11  &   20  &  22   &   85  &   87 & 27  &   52  &  54   &   117 &  119 \\ \midrule
 12  &   21  &  23   &   84  &   86 & 28  &   53  &  55   &   116 &  118 \\ \midrule
 13  &   24  &  26   &   89  &   91 & 29  &   56  &  58   &   121 &  123 \\ \midrule
 14  &   25  &  27   &   88  &   90 & 30  &   57  &  59   &   120 &  122 \\ \midrule
 15  &   28  &  30   &   93  &   95 & 31  &   60  &  62   &   125 &  127 \\ \midrule
 16  &   29  &  31   &   92  &   94 & 32  &   61  &  63   &   124 &  126 \\ 
\bottomrule
\end{tabular}
\label{tab:32MappingTable}
\end{table*}

However, on 6 core processor, cache set with different cache set index might have different mapping table. There are 2048 cache set on each set index. As a result, there exist 2048 mapping tables corresponding to 2048 set indexes. After further analysis, this article has tested every set index with 1GB physical memory (30 bits physical address). As presented in table~\ref{tab:32MappingTable}, the result shows that the mapping table of some set indexes are the same. Set index 0, 2, 65, 67 share the same mapping table. And There are 32 different mapping table. Set index ranging from 0 to 2047 fall into 32 mapping tables.
\begin{figure}[!htp]
\centering
\includegraphics[width=0.8\columnwidth]{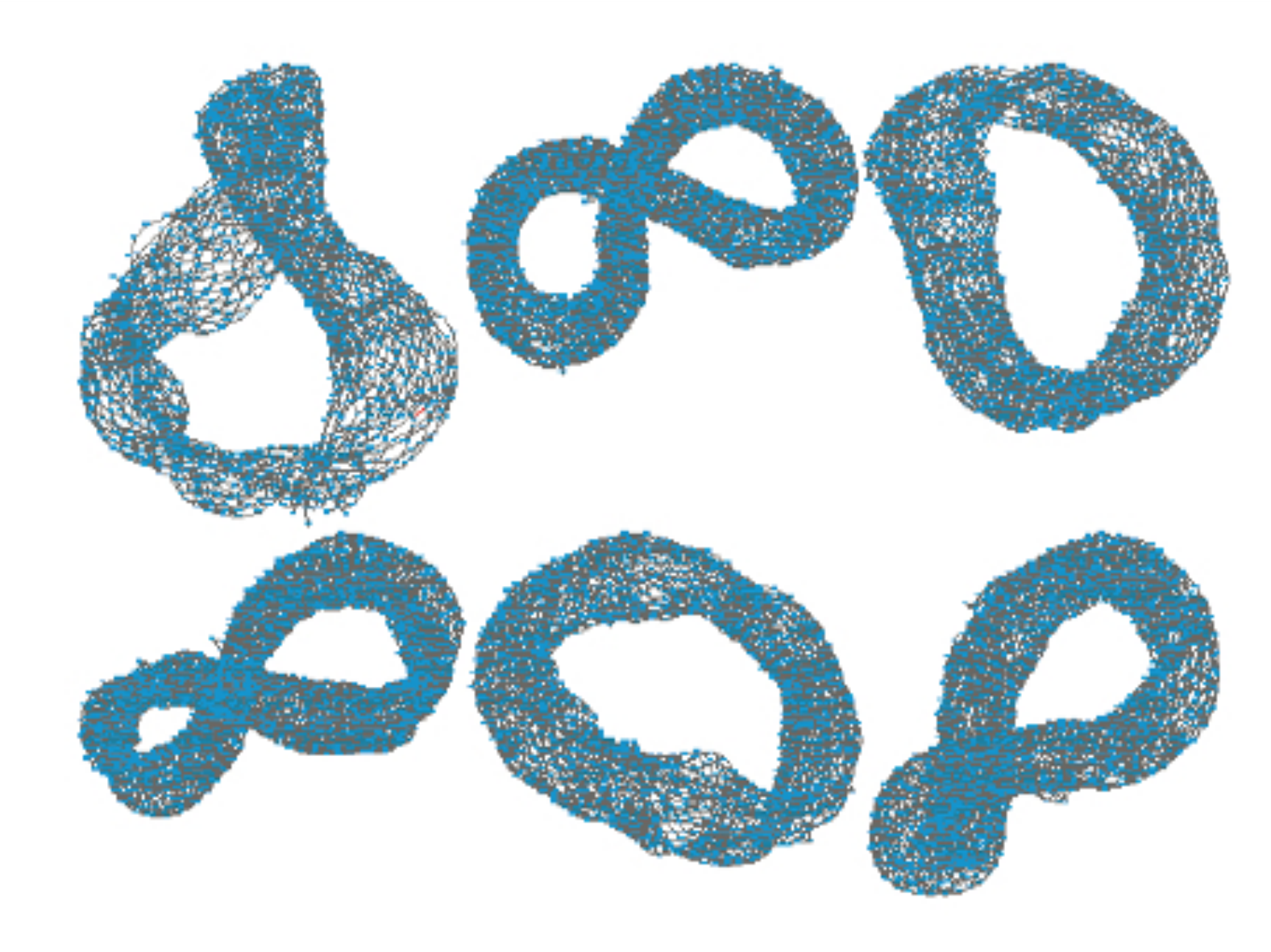}
\caption{The physical addresses sharing the same set index are divided into into six groups using Cytoscape.}
\label{fig:classification}
\end{figure}

Higher physical address bits also affect the result of hash function. On Sandy Bridge 6 core processor, this article has verified each set index with 1GB physical memory. Let $\left(a_{\PhysicalAddressWidth-1},\dots,a_0\right)$ be the binary representation of physical address, with 1GB physical memory tested, this article gets the result of bits $\left(a_{29},\dots,a_0\right)$. When it comes to the higher address bits $\left(a_{35},\dots,a_{30}\right)$, this article verifies with 64GB(36 bits physical address) physical memory. The result shows the higher bits also affects the result of hash function. However, the data blocks sharing the same set index can still be split into 6 groups. This means that substring $A_2$ is used to select the $\SliceID$ of a given $\MA$.
\section{Verification of the correctness of the cracking result of \SandyBridge cache hash function}
\subsection{Object of correctness verification}
This article proposes a method to verify the correctness of the cracking result. The cracked function will map different data blocks to different cache sets. This article verifies the correctness of the cracking result by checking that the data blocks that are indicated by the cracking result to be in one cache set are truly in one cache set.
\begin{figure}[!htbp]
\centering
\includegraphics[width=0.8\columnwidth]{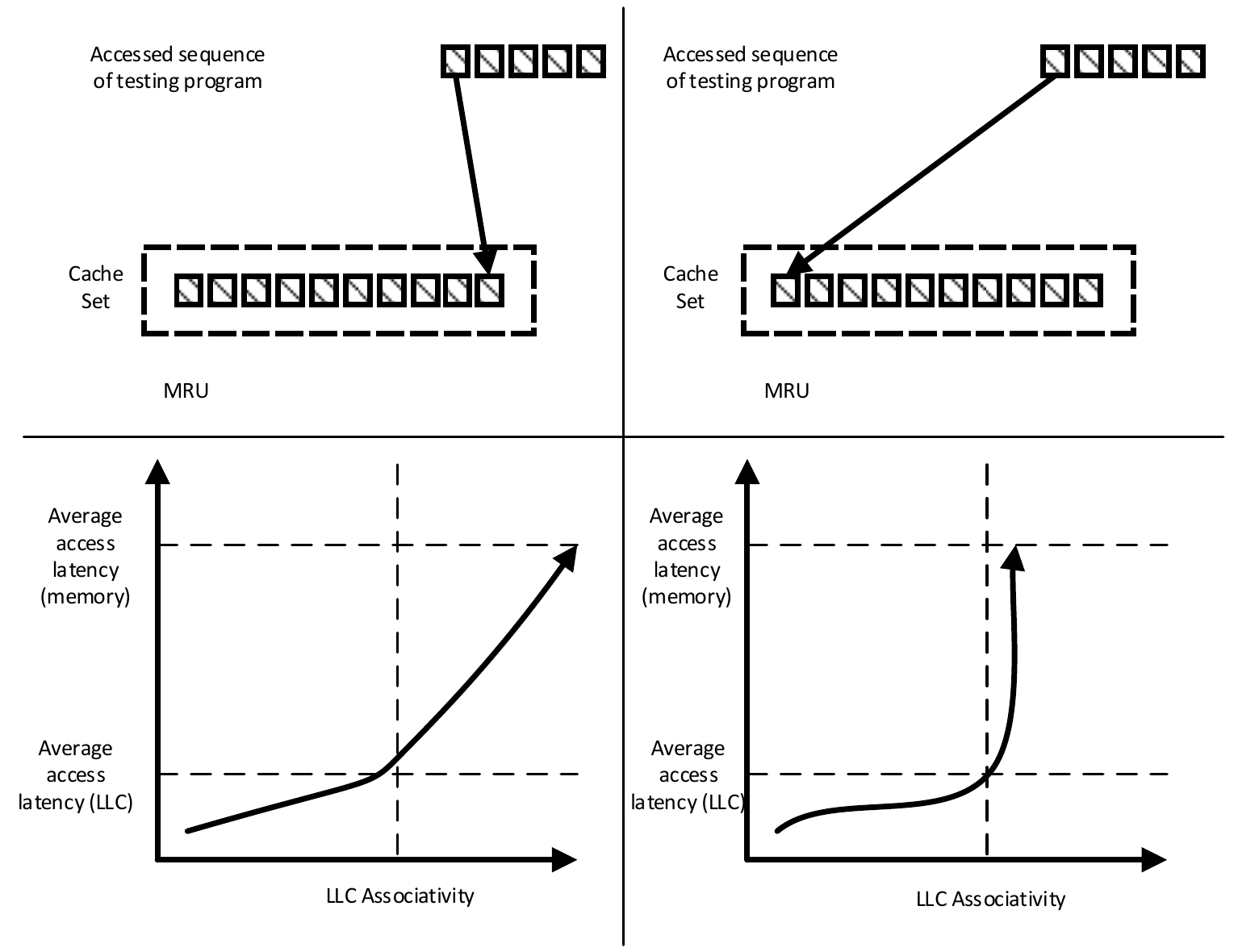}
\caption{Replacement policy affects the avearage latency.}
\label{fig:New08CacheReplacementPolicy}
\end{figure}
\begin{figure}[!htbp]
\centering
\includegraphics[width=0.8\columnwidth]{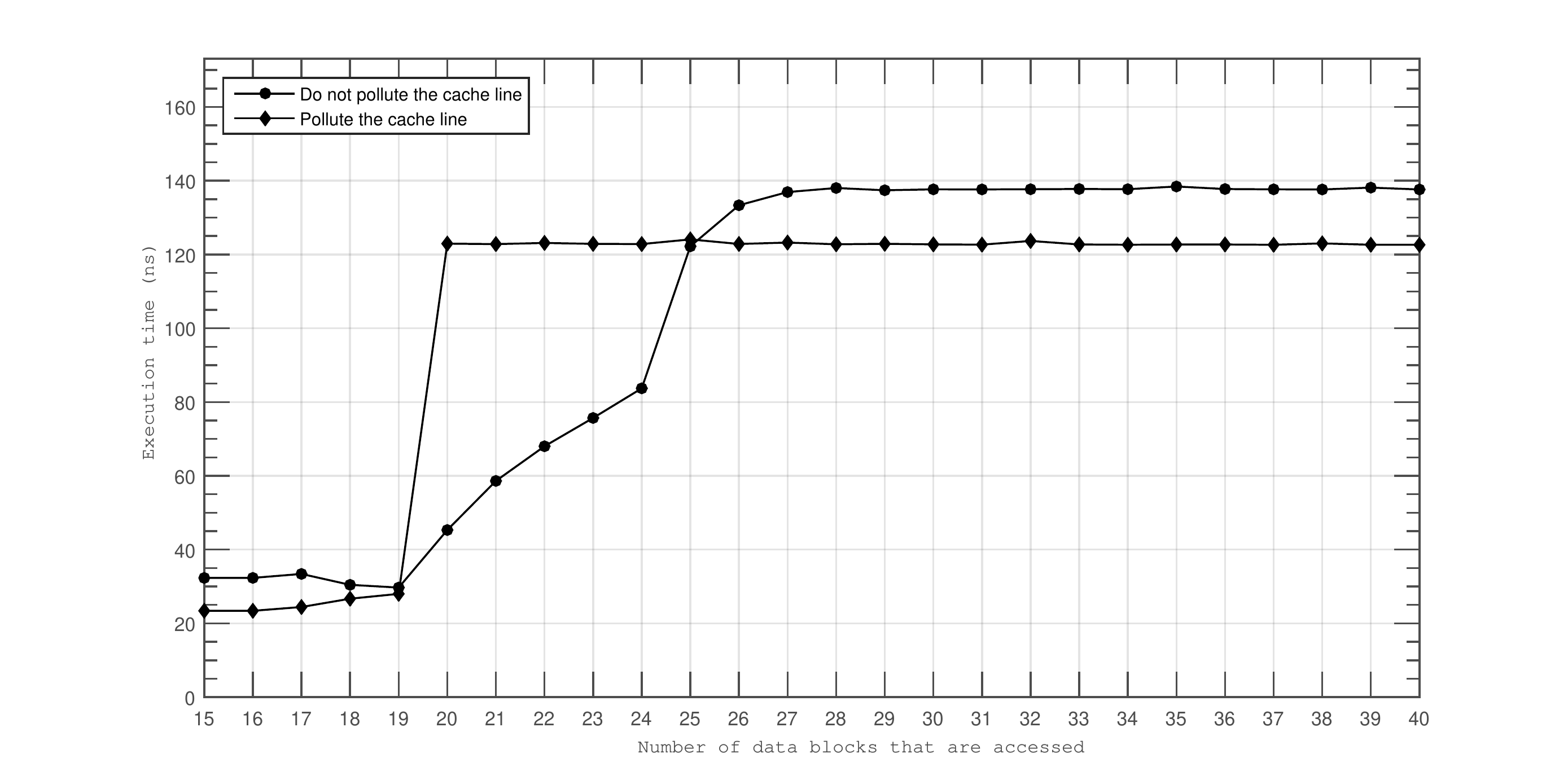}
\caption{Replacement policy affects the avearage latency.}
\label{fig:SingleThreadPolluteNopollute}
\end{figure}
\begin{figure*}[!htbp]
\centering
	\subfigure[Thread 1 accesses 1 data block.]{
		\includegraphics[width=3in]{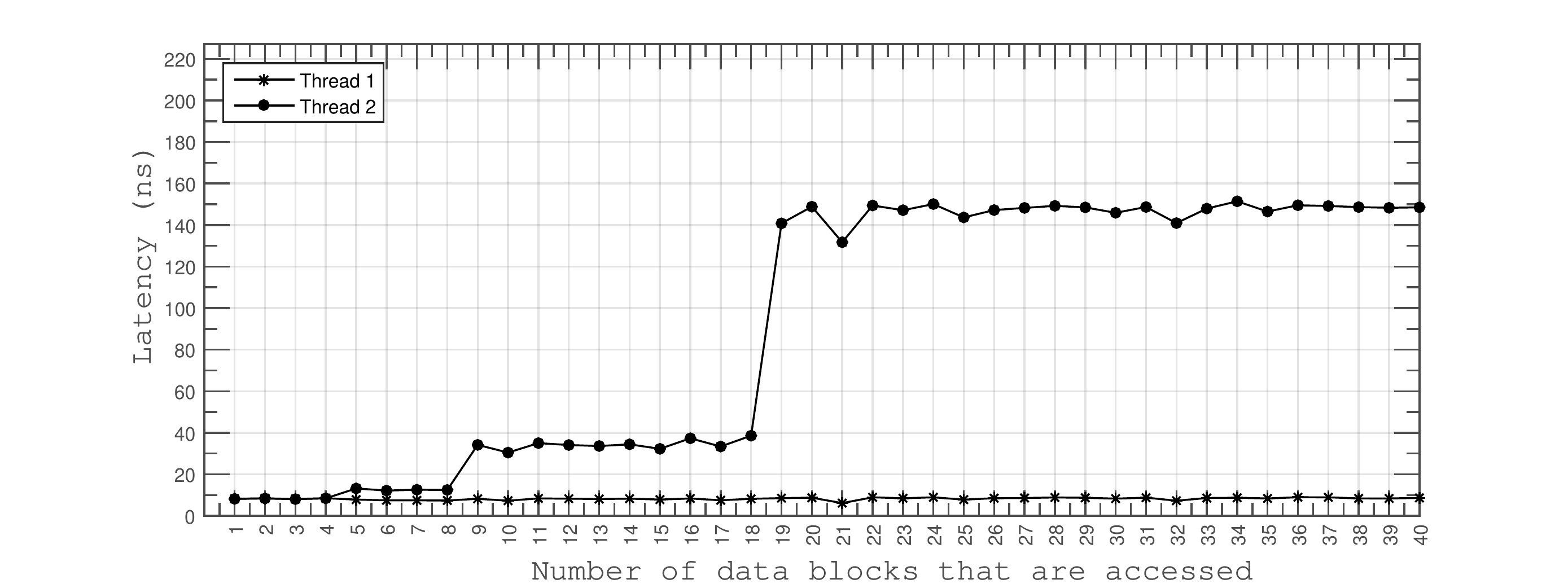}
	}
	~
	\subfigure[Thread 1 accesses 2 data blocks.]{
		\includegraphics[width=3in]{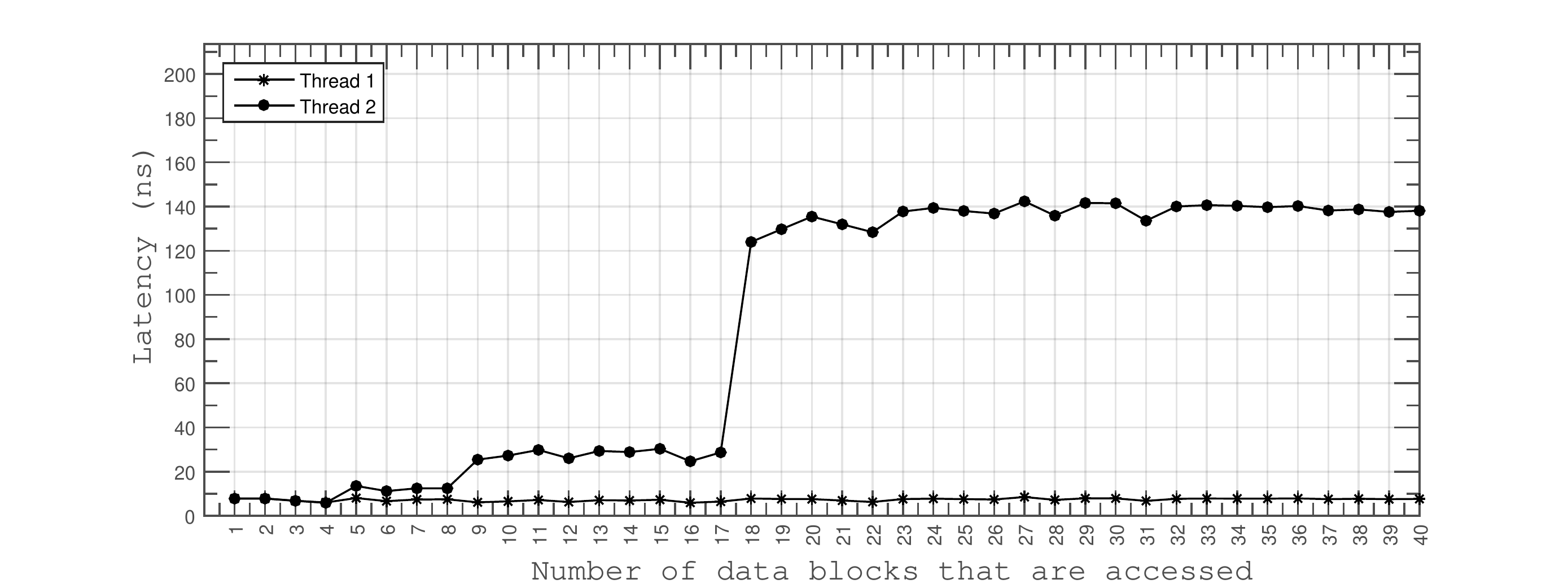}
	}
        ~
	\subfigure[Thread 1 accesses 3 data blocks.]{
		\includegraphics[width=3in]{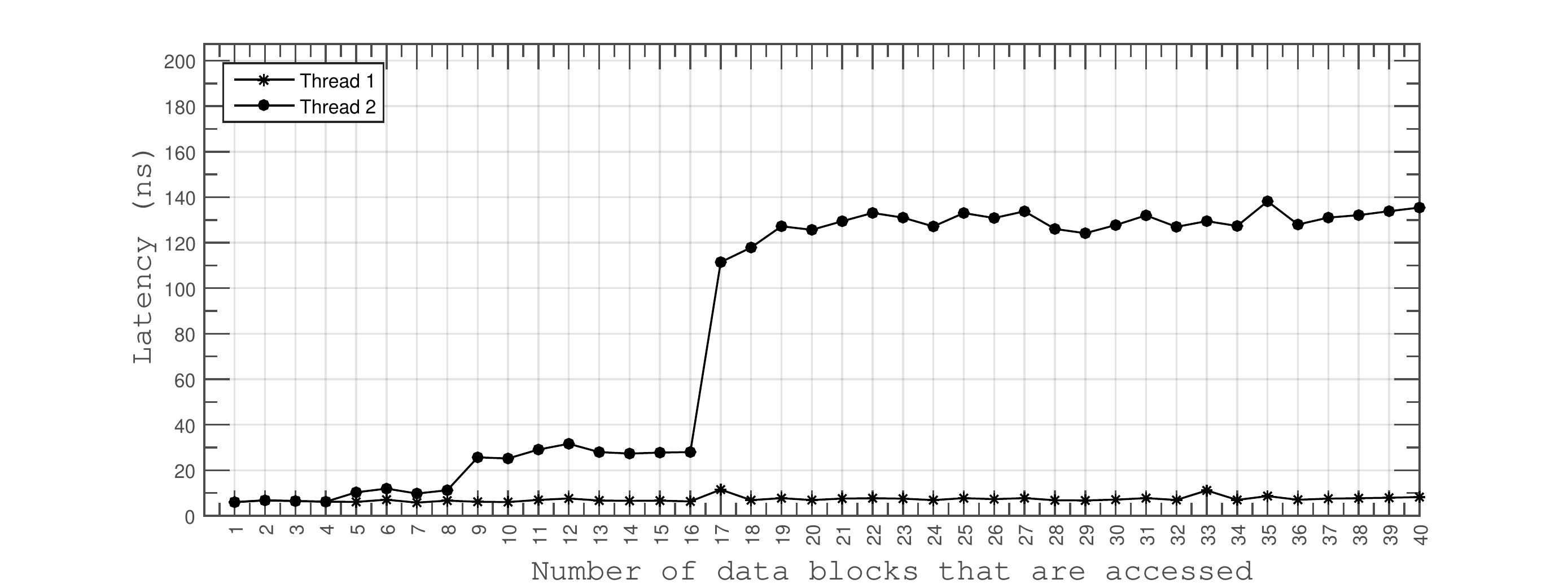}
	}
	~
	\subfigure[Thread 1 accesses 4 data blocks.]{
		\includegraphics[width=3in]{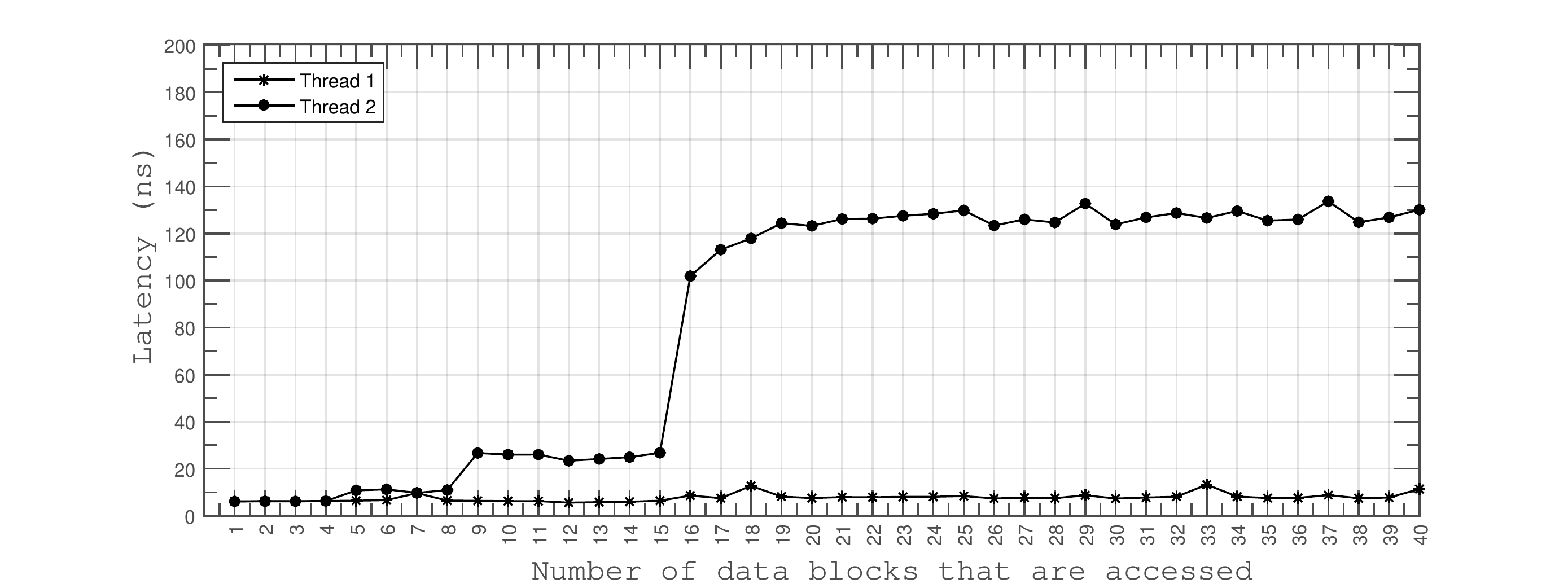}
	}
	\caption{The data blocks accessed by two threads are mapped into the same cache set.}
	\label{fig:TwoThreadsSameSets}
\end{figure*}
\begin{figure*}[!htbp]
\centering
	\subfigure[Thread 1 accesses 1 data block.]{
		\includegraphics[width=3in]{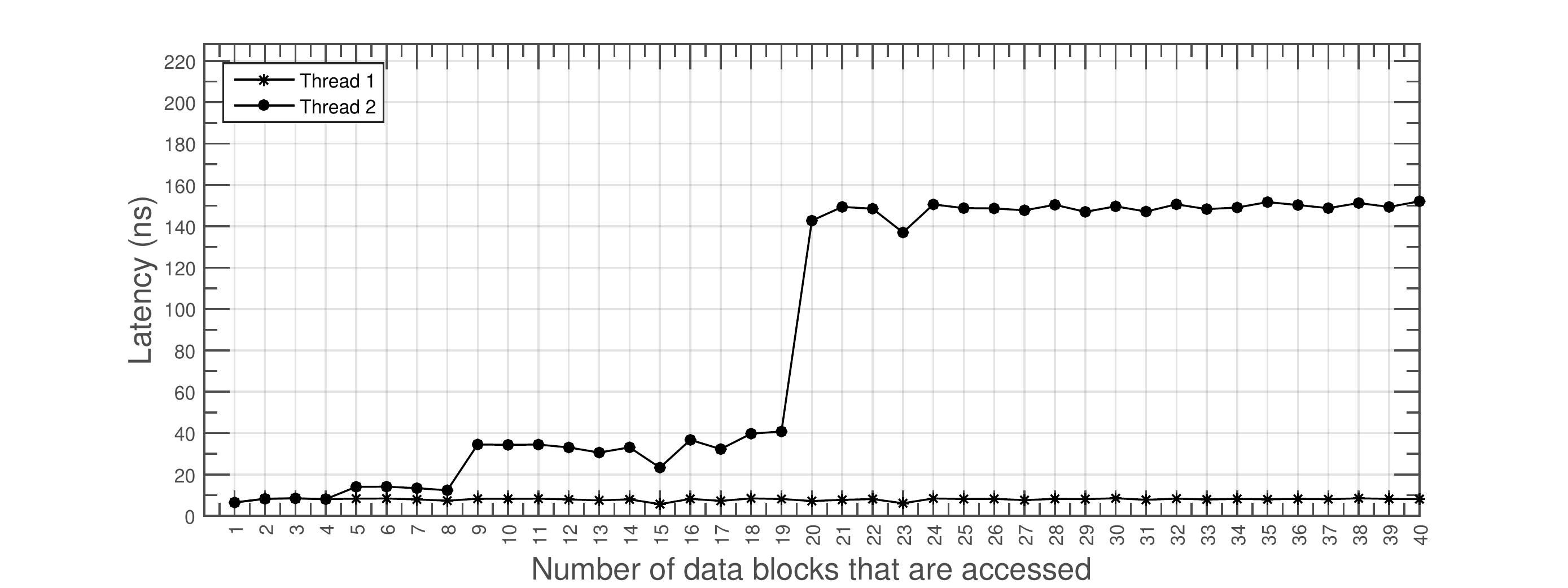}
	}
	~
	\subfigure[Thread 1 accesses 2 data blocks.]{
		\includegraphics[width=3in]{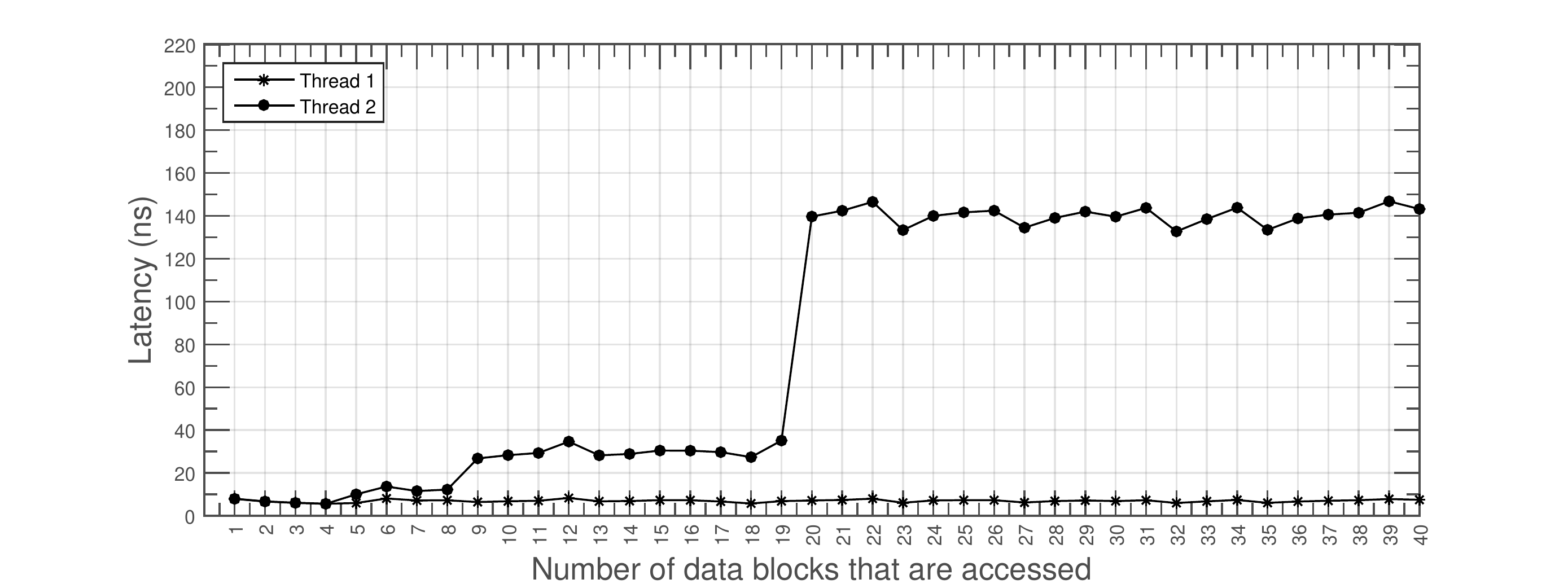}
	}
        ~
	\subfigure[Thread 1 accesses 3 data blocks.]{
		\includegraphics[width=3in]{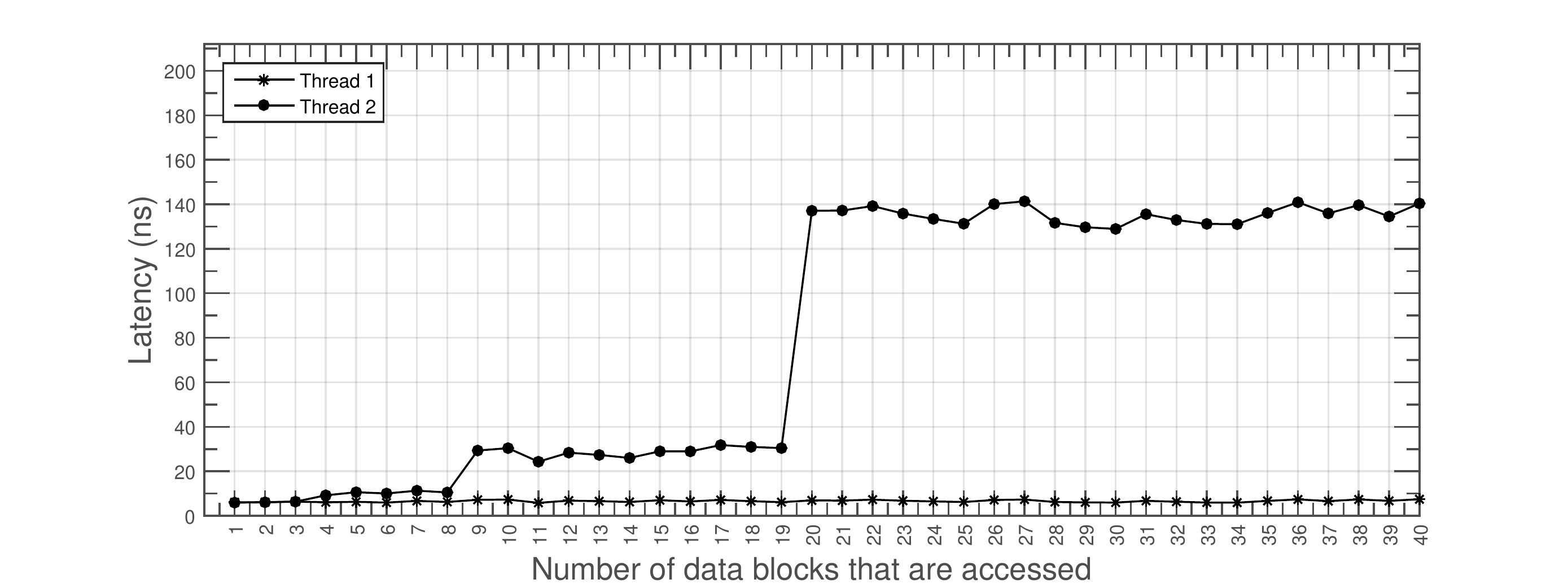}
	}
	~
	\subfigure[Thread 1 accesses 4 data blocks.]{
		\includegraphics[width=3in]{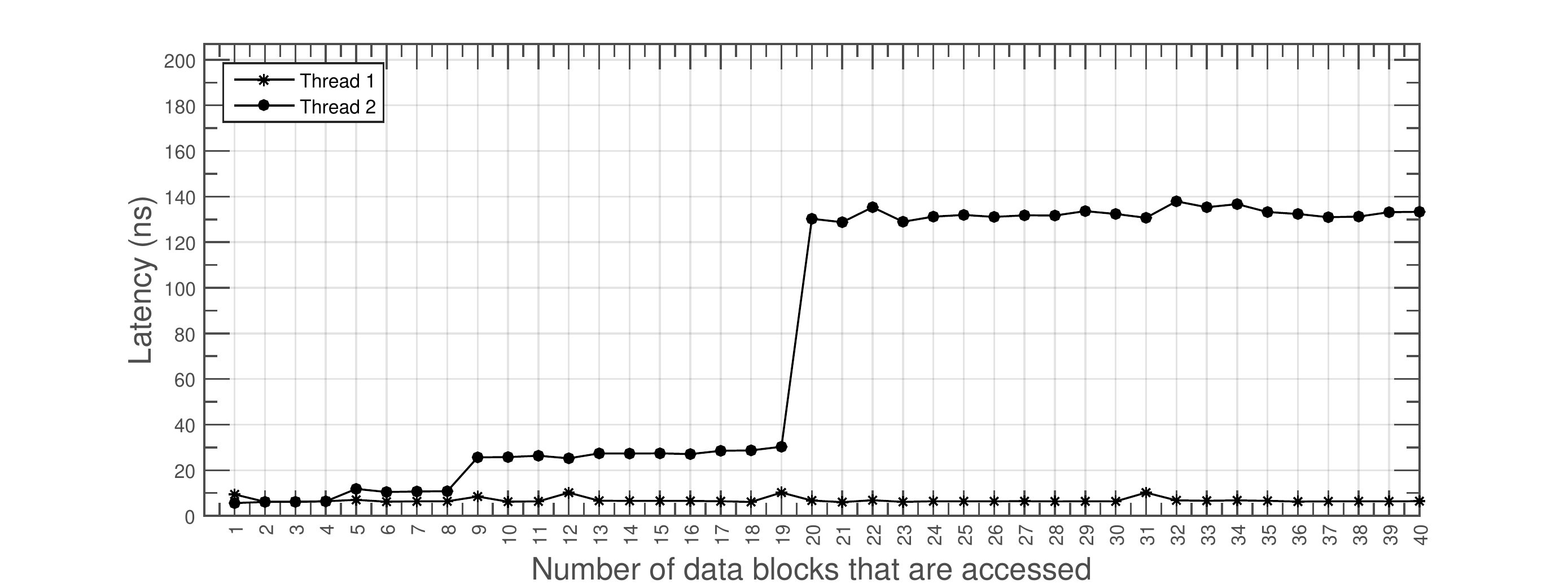}
	}
	\caption{The data blocks accessed by two threads are mapped into two different cache set.}
	\label{fig:TwoThreadsDifferentSets}
\end{figure*}

As presented in figure~\ref{fig:SingleThreadPolluteNopollute}, average access latency increases sharply when the number of data blocks accessed exceeds the associativity of LLC. This is relatively obvious. This article finds that performing another write operation after the data block is read into cache will affect cache replacement policy. As presented in figure~\ref{fig:SingleThreadPolluteNopollute}, polluting the cache line means that adding another write operation, and do not pollute the cache line means perform only dependent read operation. It can be seen from the figure that performing another read operation causes the average access latency to increase slowly compared to the read operation only configuration. One possible explanation is as shown figure~\ref{fig:New08CacheReplacementPolicy}, when a new cache line arrives, if the newly accessed data block is inserted into the MRU position, when the number of data blocks accessed exceeds the associativity of LLC, even if only exceeds by one, the average access latency will increase sharply. The method is able to check every cache set.

Verify the correctness of the cracking result using two threads. In this verifying scenario, use two threads to perform data dependent access. The average access latency is recorded for each configuration. In the first configuration, the data blocks accessed by thread 1 and the data blocks accessed by thread 2 are mapped to different cache sets, the results is presented in figure~\ref{fig:TwoThreadsDifferentSets}. In the second configuration, the data blocks accessed by thread 1 and the data blocks accessed by thread 2 are mapped to different cache sets, the results is presented in figure~\ref{fig:TwoThreadsSameSets}. For both configurations, the number of data blocks accessed by thread 2 vary from 1 to 40. And the number of data block accessed by thread 1 varies from 1 to 4. When the number of data blocks accessed by thread 1 exceeds the associativity, average latency will increase sharply. When thread 1 performs data dependent access on 1 data block, in the first configuration, as the data blocks accessed by two threads are mapped to the same cache set and thread 1 accessed 1 data block, average access latency of thread 2 increases sharply when the number of accessed data blocks of thread 2 is 18; On the contrary, when the data blocks accessed by two threads are mapped to different cache sets, average access latency of thread 2 increases sharply when the number of accessed data blocks of thread 2 is 18. This article gets similar results when the number of data blocks access by thread1 varies from 1 to 4.

Write operation has effect on cache replacement policy. In the following test, the program performs data dependent access, two configuration of the program is as follows:
\begin{inparaenum}[(1)]
\item when the data block is read, perform a write operation to make the cache line dirty;
\item do not make the cache line dirty.
\end{inparaenum}

Let $L_{average}$ be the average access latency, $L_{memory}$ be the access latency of memory, $L_{LLC}$ be the access latency of LLC, $N$ be the associativity of LLC, $n$ be the number of data blocks that are accessed sequentially by the testing program, then the relationship between average access latency and the number of data blocks accessed can be describes as:
\begin{equation}   
	L_{average}=
      \begin{cases} 
	      L_{LLC}, & \text{if}\ n < N \\
	      L_{memory}(\frac{n}{N} - 1) + L_{LLC}, & \text{if}\ n \ge N 
      \end{cases}
      \label{Equ:slow}
\end{equation}
\begin{figure*}
\centering
	\subfigure[18 addresses]{
		\includegraphics[width=2.5in]{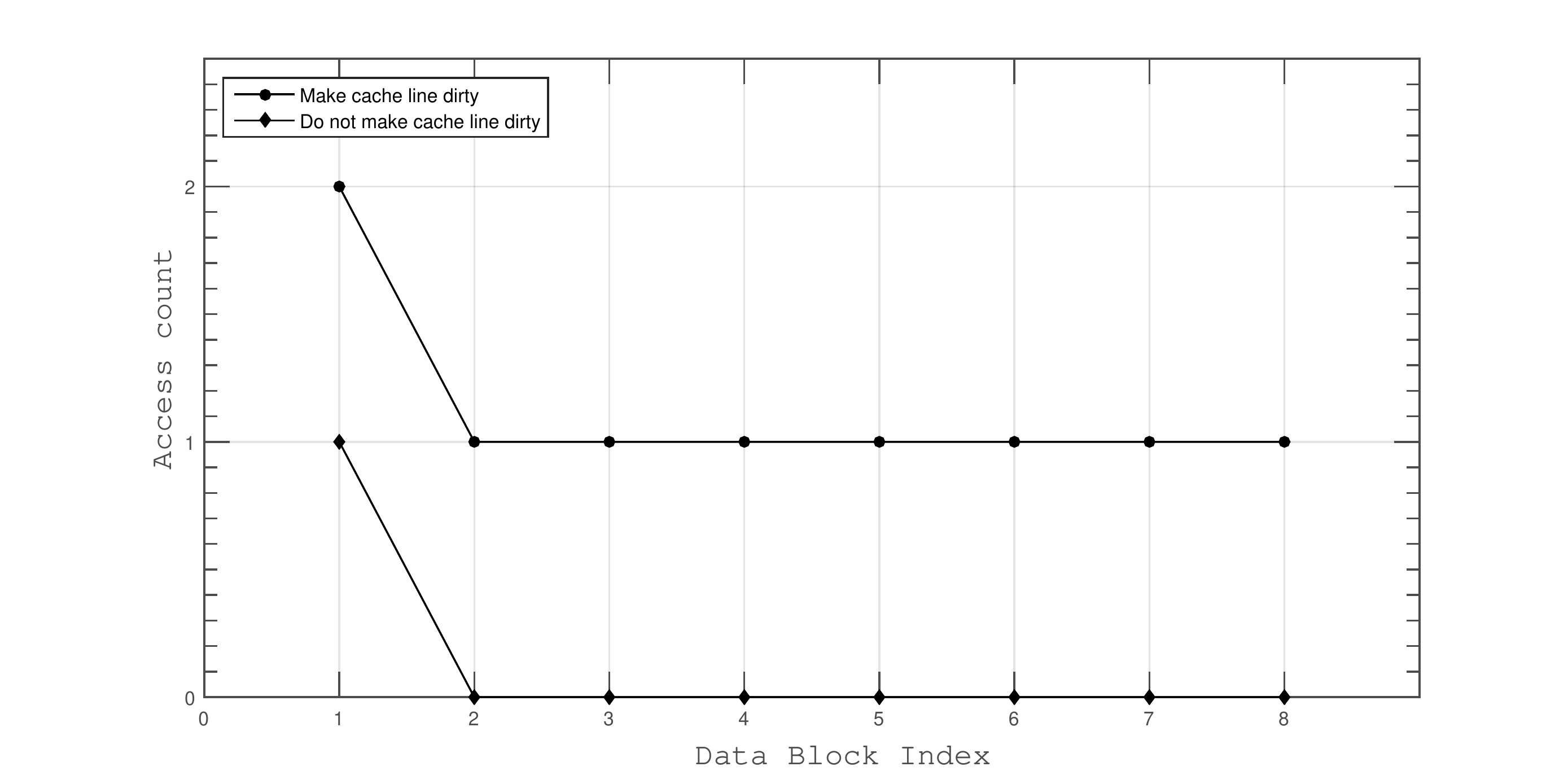}
	}
	~
	\subfigure[19 addresses]{
		\includegraphics[width=2.5in]{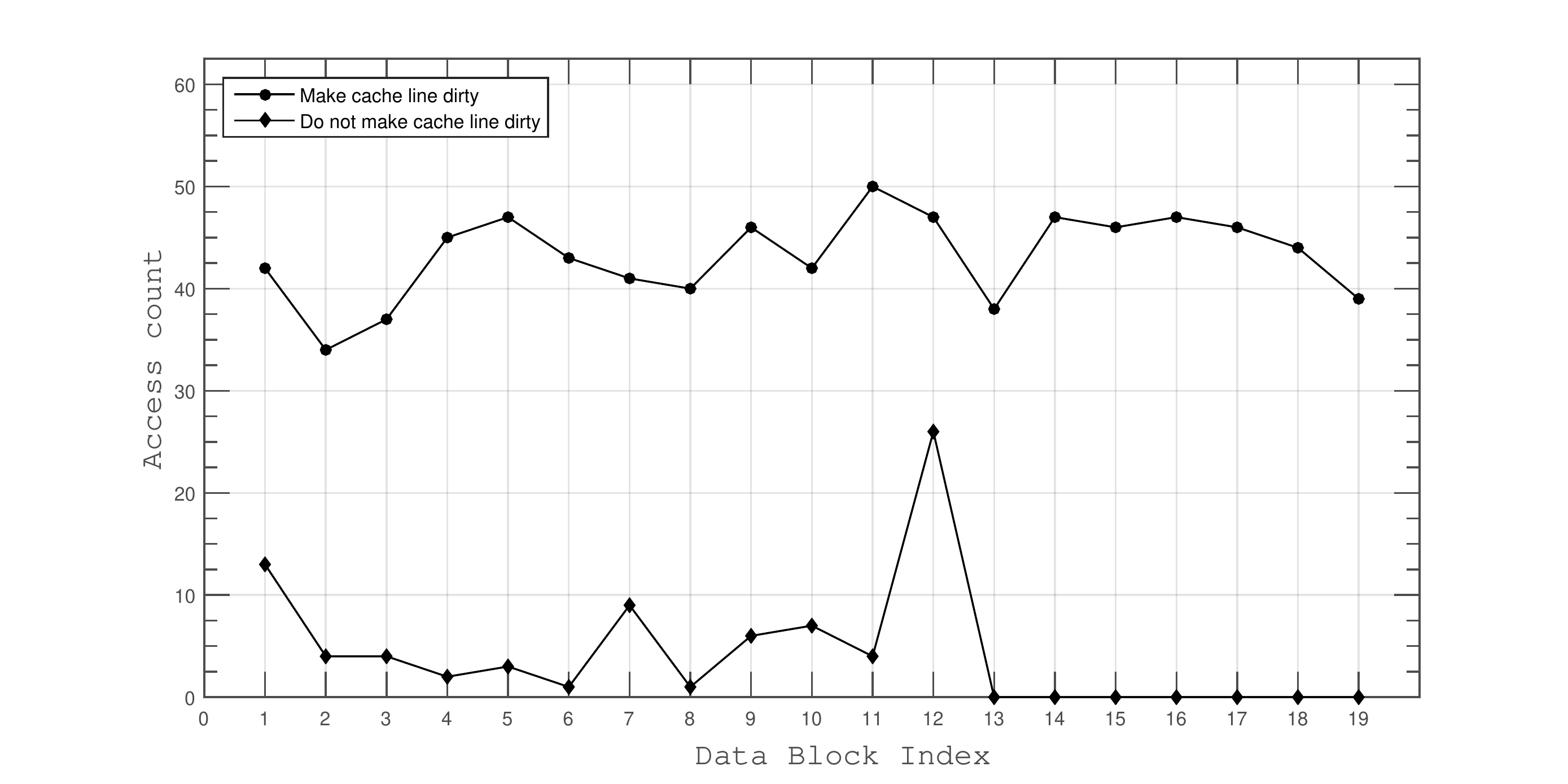}
	}
	~
	\subfigure[20 addresses]{
		\includegraphics[width=2.5in]{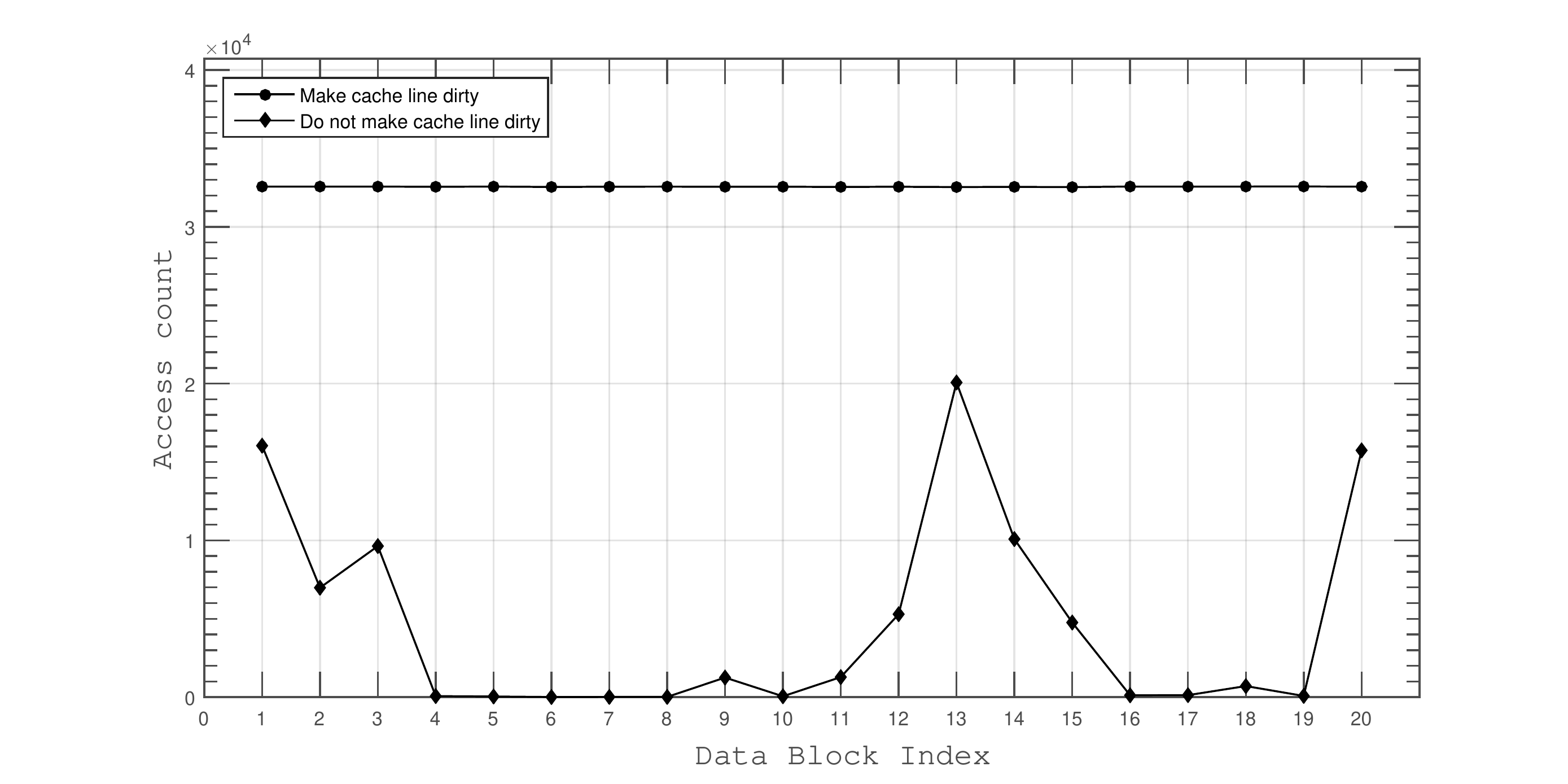}
	}
	~
	\subfigure[21 addresses]{
		\includegraphics[width=2.5in]{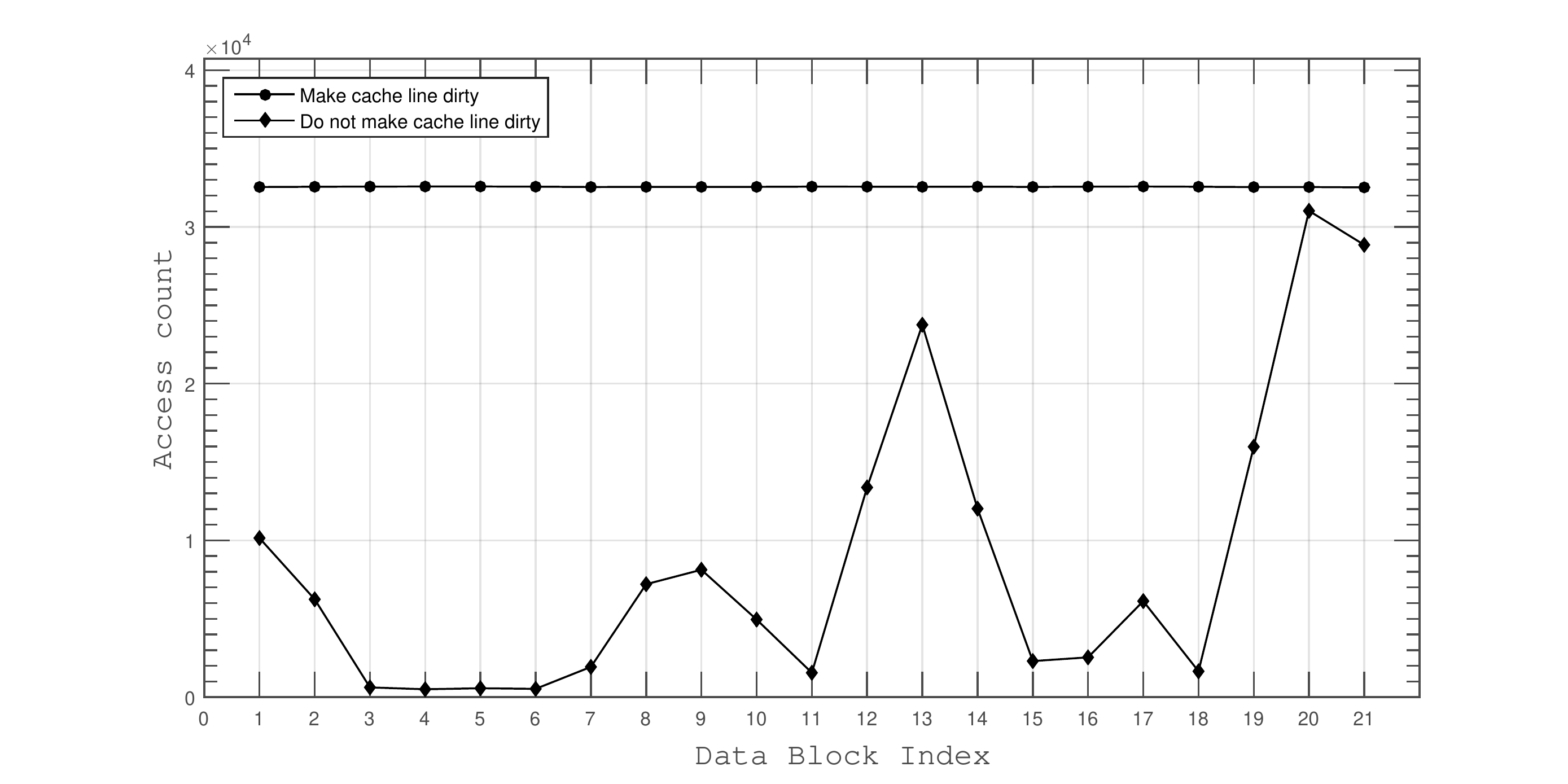}
	}
	~
	\subfigure[22 addresses]{
		\includegraphics[width=2.5in]{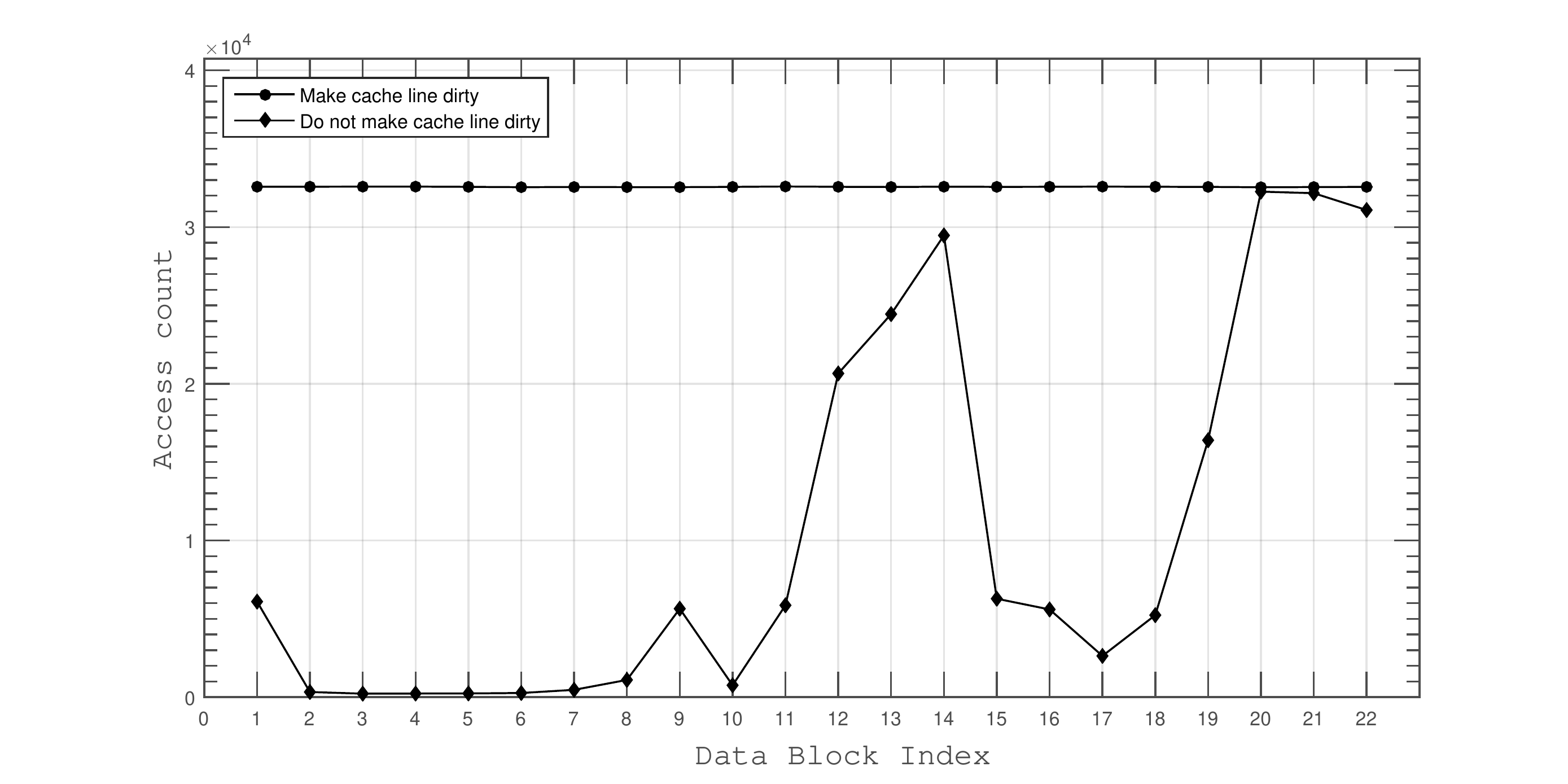}
	}
	~

	\subfigure[23 addresses]{
		\includegraphics[width=2.5in]{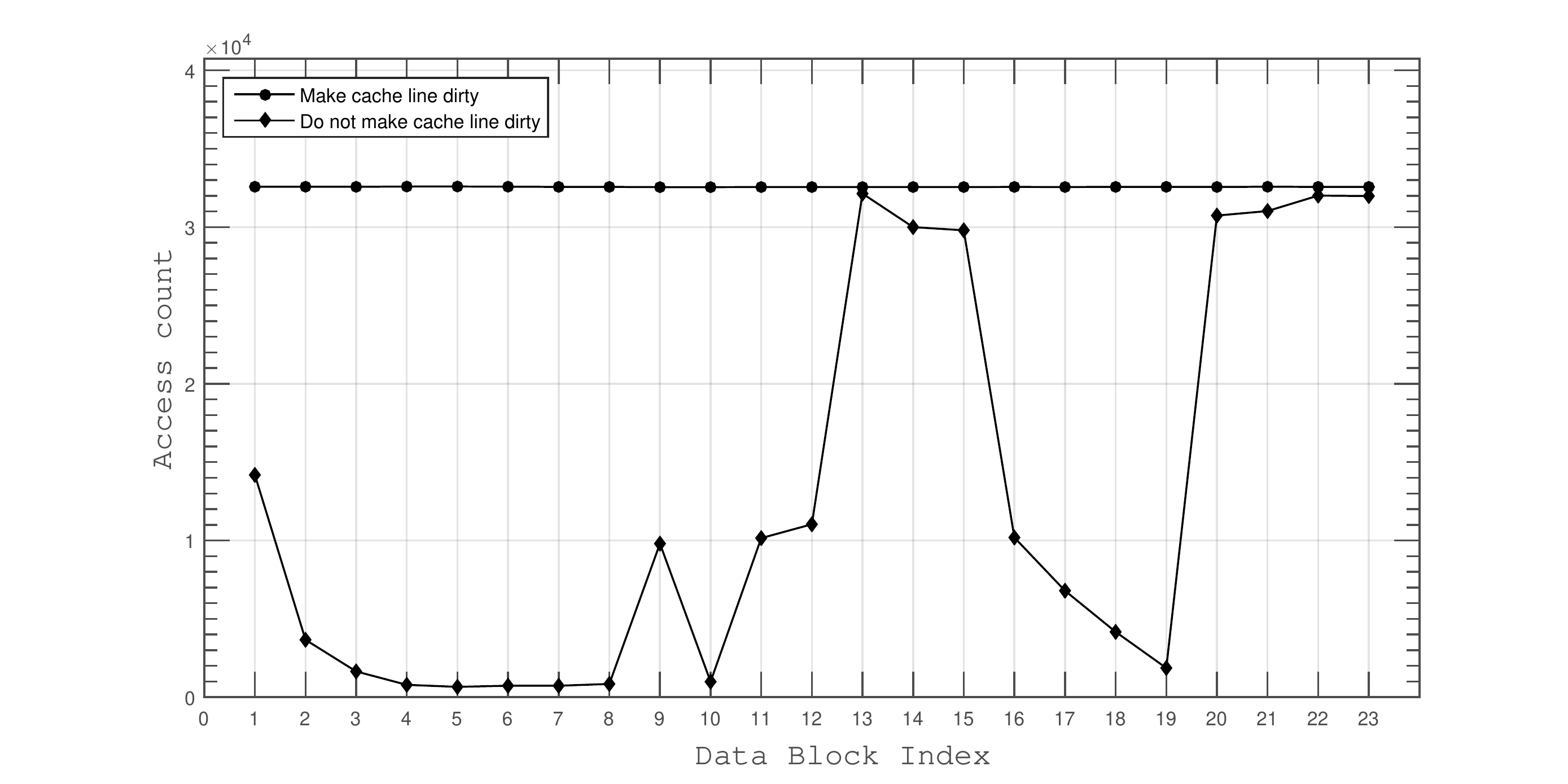}
	}
	~
	\subfigure[24 addresses]{
		\includegraphics[width=2.5in]{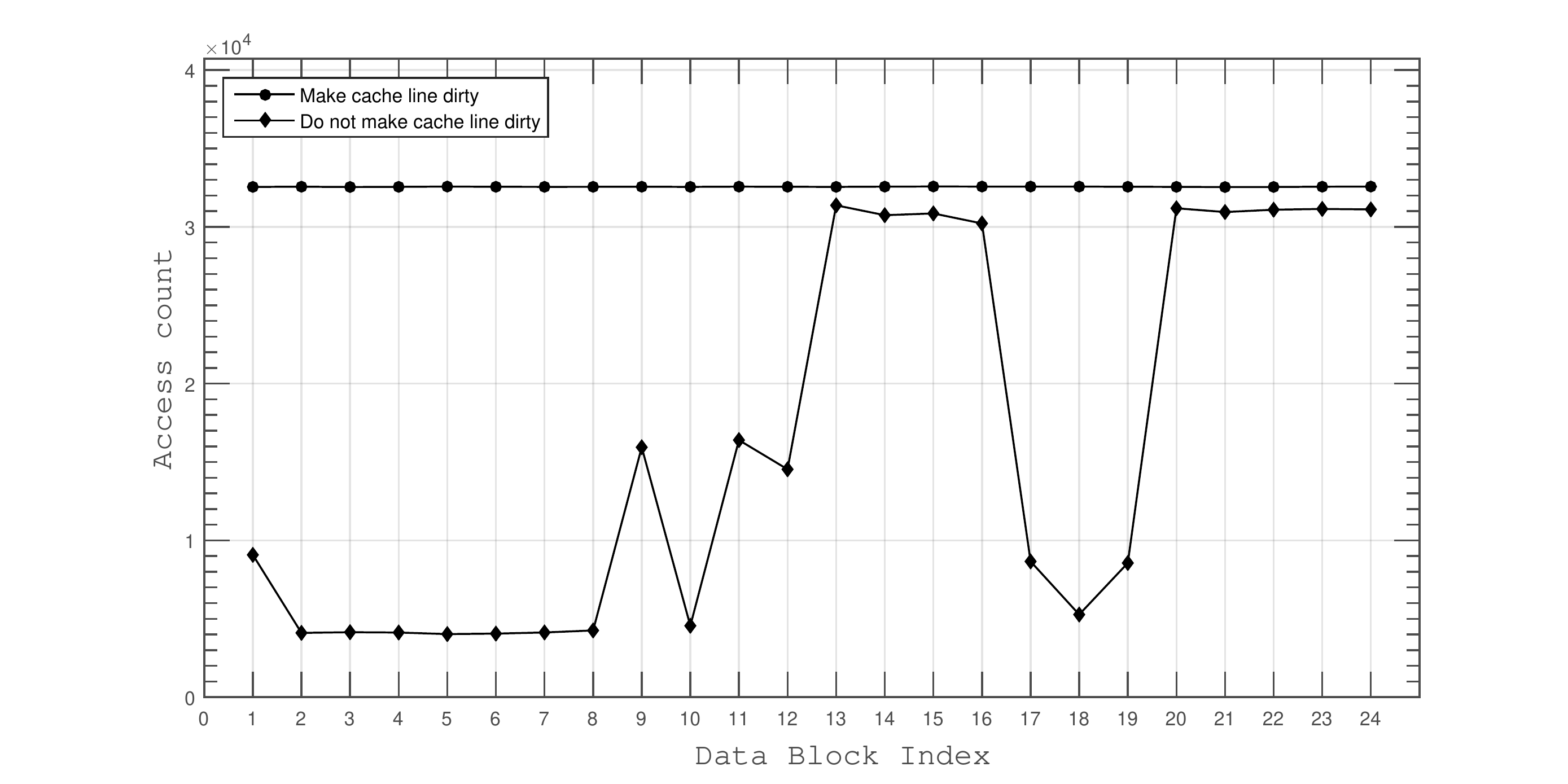}
	}
	\caption{When test program performs data dependent access of different number of data blocks, the access number distribution of different data blocks in a period of time.}
	\label{fig:pollute_nopollute}
\end{figure*}
Memory reference trace collected with HMTT offers a different prospective on cache replacement policy. This article further analyzes the trace: cut a segment of the whole trace, and count how many times each data block is accessed in this segment of trace (or this period of execution). As presented in figure~\ref{fig:pollute_nopollute}, without polluting the cache line, for simplicity, label each data block accessed with an index, the access number of each data block is uniform. However, when perform an extra write operation to pollute the cache line, the access number of different data blocks becomes non-uniform, which means that some of the data blocks are held in cache longer than the other part of data blocks. This fact reveals the fact that write operation affects the cache replacement policy. 
\section{ulcc}
User Level Cache Control (ULCC)~\cite{ding2011ulcc} is a software package to implement cache partition using page coloring. It improves performance of multi-threaded program by enforcing a user demand cache capacity allocation. By modifying the macro that extracts page color from physical address, this article ported ULCC to Intel Sandy processor.
\begin{figure}[!htbp]
\centering
\includegraphics[width=0.8\columnwidth]{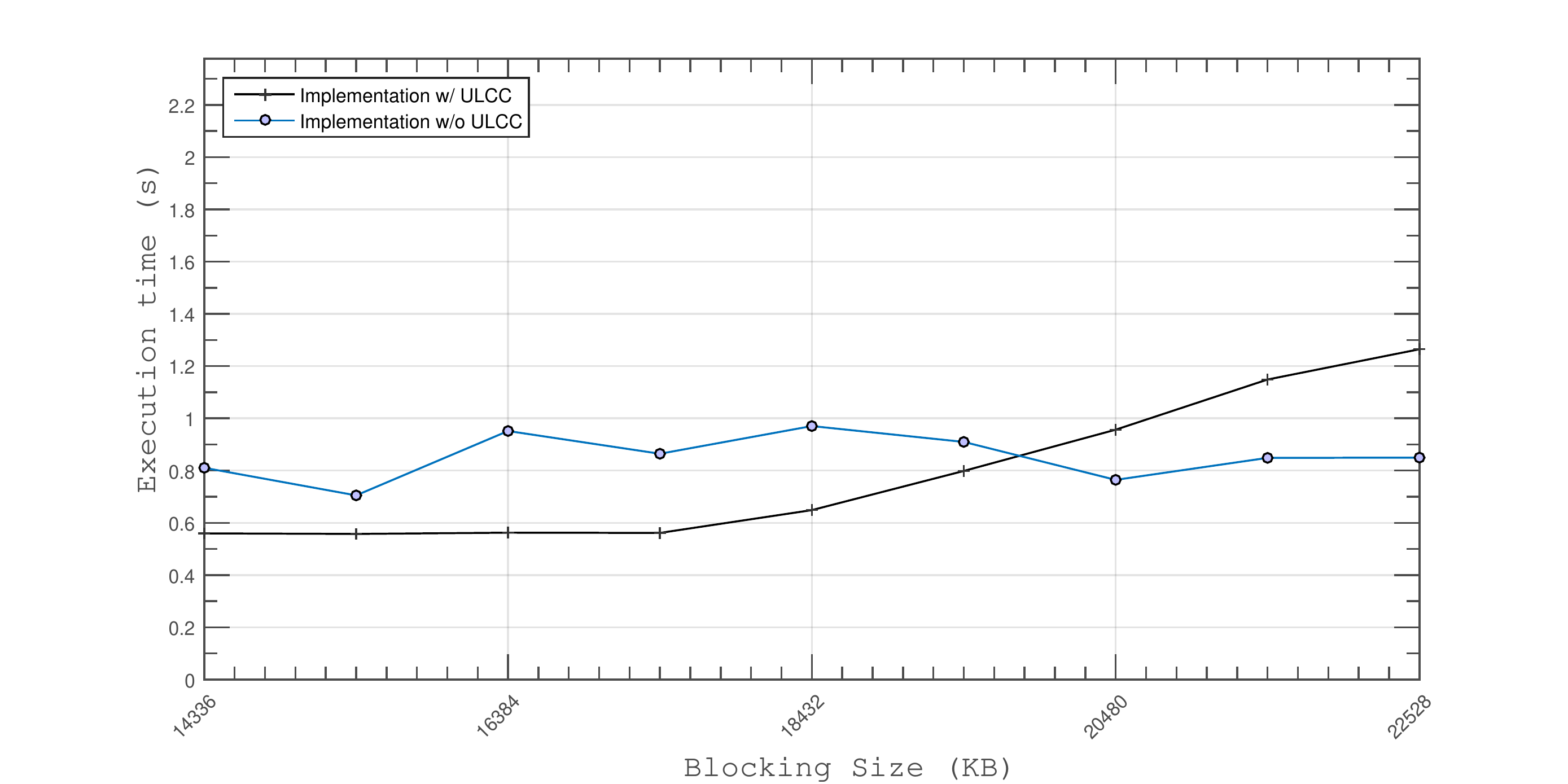}
\caption{The execution time of MergeSort implementations with and without ULCC support.}
\label{fig:ULCCSortBlockTime}
\end{figure}
\subsection{MergeSort}
%
MergeSort is implemented in multiple threads. During the execution of program, the intermediate result of every block is highly reused. ULCC improves the performance by allocating different cache capacity for data of different reuse degree.

The performance of the program with and without ULCC support is depicted in figure~\ref{fig:ULCCSortBlockTime}. When choosing size of sorting block properly, the execution time is reduced by 20\%. The result is using one thread to finish merge sort. So the performance gain is only from preventing cache pollution of data in one thread.
\subsection{MatMul}
The MatMul program multiplies two double precision matrices A and B, and produces the product matrix C. To achieve necessary data reuse in LLC, the matrix multiplication is carried out block by block. For the block a on the $i$th block row and $j$th block column of matrix A, it is multiplied with all the blocks on the $j$th block row of matrix B, and the results are accumulated into the blocks on the $i$th block row of matrix C. So the data in matrix A is of high reuse degree and before the program finishes the computation with block a, it is desirable that the data in a can be kept in the cache. However, without a dedicated space for block a, the data in it may be repeatedly evicted from the cache before its next use every time the program switches blocks in matrix B and matrix C, even with a rather small block size. To reduce the chance that the data in each block of matrix A is evicted from the last level cache prematurely, the size of sorting block should has a most suitable size.

\begin{figure}[!htp]
\centering
\includegraphics[width=0.8\columnwidth]{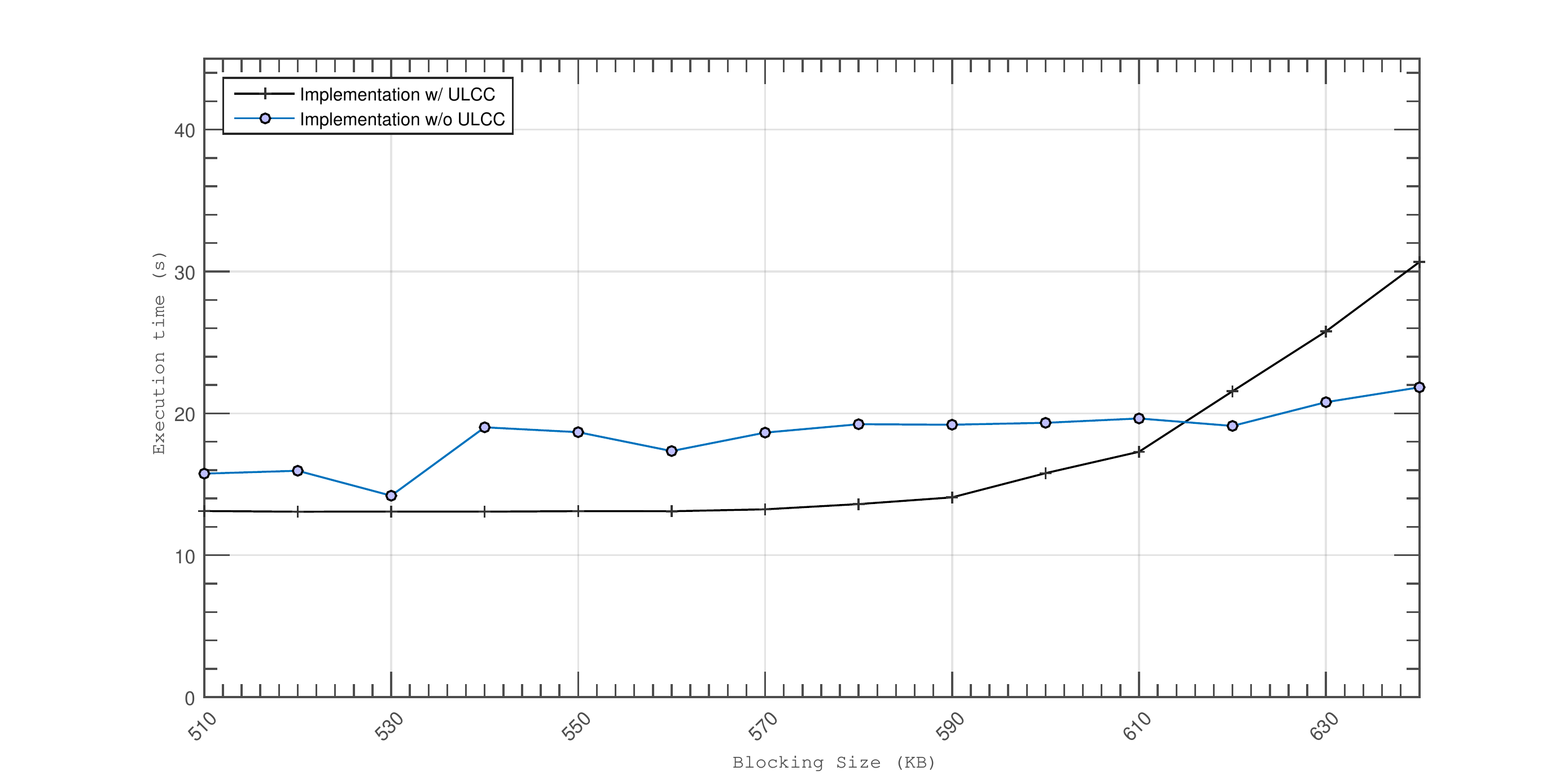}
\caption{The execution time of MatMul implementations with and without ULCC support.}
\label{fig:MMBlockTime}
\end{figure}

The performance of the program with and without ULCC support is depicted in figure~\ref{fig:MMBlockTime}. When choose data block element properly to make sure that the frequently used data can be held in cache, this article achieves the same performance improvement as presented in~\cite{ding2011ulcc}. This further proves the correctness of the cracking results.
\section{General test}
This article proposed a method to crack hash function without the support of HMTT. The core of idea is that when the number of accessed addresses exceeds the associativity of cache set, the average access latency will increase sharply.

The main procedures of the cracking method without support of HMTT is as follows:
\begin{inparaenum}[(1)]
\item Verify that some bits of physical address are used directly to select cache set; Divide all data blocks into different groups based on set index;
\item  choose \Associativity data blocks that are mapped to one cache set from the data blocks sharing the same cache set index,; Access these data blocks sequentially, the average access latency will be close to the access latency of main memory; Put these data blocks in the classified group of data blocks;
\item Choose $\Associativity - 1$ data blocks from the classified group and choose one data block from the unclassified group of data blocks, perform data dependent access of these data blocks, there are two results: the first is that the average access latency is close to the latency of main memory, and this means that the one data block chosen from the unclassified group of data blocks is mapped to the same cache set with the other $\Associativity - 1$ data blocks; the second is that the average access latency is close to the latency of LLC, and this means that the one data block chosen from the unclassified group is not mapped to the same cache set with the other $\Associativity - 1$ data blocks.
\item Perform step 2 and step 3 until all data blocks sharing the same set index have been labeled as classified or unclassified.
\item Choose another data block from the data blocks which are labeled as unclassified, start again from step 1;	
\end{inparaenum}

The problem with this method is that it takes too long to finish the test. 
\section{Conclusions}
On Intel Sandy Bridge processor, last level cache (LLC) is divided into cache slices and all physical addresses are distributed across all cache slices using a hash function. With this undocumented hash function existing, it is impossible to implement cache partition based on page coloring.

This article cracks the hash function on two types of Intel Sandy Bridge processors. It's true on both 4 core and 6 core processors that bit substring of physical address is used to select cache sets. What is different is that: on Intel Sandy 4 core processor, the mapping relationship for different cache set indexes is the same. And on Intel Sandy Bridge 4 core processor, the cracked hash function is reduced to a simple formula. On the contrary, on Intel Sandy Bridge 6 core processor, different cache set indexes have different mapping relationship. The article has not reduced the hash function to a simple formula. Instead, the hash function is presented in the form of mapping tables.

This article proves that it's possible to implement cache partition based on page coloring. On 4 core processor, based on the cracking result, it's easy to implement cache partition based on page coloring. On 6 core processor, without reducing the hash function to a simple formula, cache partition can at least be implemented based on set index, as bit substring of physical address is used to select cache sets.

\bibliographystyle{abbrv}
\bibliography{IEEEexample}

\end{document}